\documentclass[aps,prd,epsf,nofootinbib]{revtex4-2}

\usepackage{graphicx}
\usepackage{dcolumn}
\usepackage{bm}

\usepackage{amsmath}
\usepackage{subfigure}
\usepackage{subfig}
\usepackage{xcolor}
\usepackage{booktabs}
\usepackage{topcapt}
\usepackage{subfiles}
\usepackage{color}

\usepackage[normalem]{ulem}

\definecolor{amethyst}{rgb}{0.6, 0.4, 0.8}

\usepackage{calrsfs}
\DeclareMathAlphabet{\pazocal}{OMS}{zplm}{m}{n}
\newcommand{\Ld}{\mathcal{D}}

\begin{document}

\title{$H(4)$ tensor representations for the lattice Landau gauge gluon propagator and the estimation of lattice artefacts}

\author{Guilherme T. R. Catumba}
\email{guitelo96@gmail.com}

\author{Orlando Oliveira}
\email{orlando@uc.pt}

\author{Paulo J. Silva}
\email{psilva@uc.pt}

\affiliation{CFisUC, Departamento de F\'{\i}sica, Universidade de Coimbra, 3004-516 Coimbra, Portugal
}%

\date{\today}

\begin{abstract}
The use of lattice tensor representations is explored to investigate the lattice Landau gauge gluon propagator for the pure SU(3) 
Yang-Mills gauge theory in 4D.  The analysis of several tensor bases allows to quantify the completeness of the various tensor bases
considered, the deviations of the lattice results from the continuum theory due to the lattice artefacts and estimate the theoretical uncertainty 
in the propagator. Furthermore, our analysis tests continuum based relations with the lattice data and show that the lattice Landau gauge gluon 
propagator is described by a unique form factor, as in the continuum formulation.
\end{abstract}

\maketitle
\tableofcontents
\section{Introduction and Motivation}

The  gluon propagator is a fundamental two point, gauge dependent, QCD Green function that has been thoroughly studied in the
Landau gauge using non-perturbative techniques to solve the theory, namely lattice QCD simulations 
\cite{Leinweber:1998uu,Becirevic:1999uc,Becirevic:1999hj,Boucaud:2000nd,Silva:2004bv,Sternbeck:2005tk,Boucaud:2005xn,Cucchieri:2007zm,Cucchieri:2007md,Cucchieri:2007rg,Oliveira:2007px,Maas:2008ri,Cucchieri:2009kk,Bogolubsky:2009dc,Cucchieri:2009zt,Dudal:2010tf,Oliveira:2010xc,RodriguezQuintero:2010wy,Ilgenfritz:2010gu,Maas:2011ej,Cucchieri:2011ig,Cucchieri:2012ii,Oliveira:2012eh,Aouane:2012bk,Ayala:2012pb,Dudal:2013yva,Maas:2014tza,Maas:2015nva,Bicudo:2015rma,Cucchieri:2016qyc,Silva:2016onh,Cucchieri:2016jwg,Duarte:2016iko,Athenodorou:2016gsa,Oliveira:2016stx,Boucaud:2017ksi,Duarte:2017wte,Maas:2017csm,Boucaud:2018xup,Dudal:2018cli,Cui:2019dwv,Li:2019hyv,Dudal:2019gvn}
and continuum based approaches \cite{Cornwall:1981zr,Aguilar:2001zy,Alkofer:2003vj,Alkofer:2003jr,Alkofer:2003jj,Fischer:2005ui,Aguilar:2006gr,Shibata:2007eq,Huber:2007kc,Boucaud:2007va,Boucaud:2007hy,Dudal:2007cw,Dudal:2008sp,Fischer:2008uz,Aguilar:2008xm,Dudal:2008rm,Aguilar:2009ke,Tissier:2010ts,Aguilar:2010zx,Dudal:2011gd,Tissier:2011ey,Kondo:2011ab,Serreau:2012cg,Strauss:2012dg,Aguilar:2012rz,Pelaez:2014mxa,Cornwall:2013zra,Serreau:2013ila,Aguilar:2014tka,Siringo:2014lva,Serreau:2015yna,Huber:2015ria,Cornwall:2015lna,Reinosa:2015gxn,Frasca:2015yva,Dudal:2015khv,Aguilar:2016vin,Aguilar:2016ock,Siringo:2015wtx,Siringo:2016jrc,Lowdon:2015fig,Pelaez:2017bhh,Reinosa:2017qtf,Cyrol:2017ewj,Lowdon:2017uqe,Tissier:2017fqf,Chaichian:2018cyv,Siringo:2018uho,Kondo:2018qus,Lowdon:2018mbn,Siringo:2019qwx,Maas:2019ggf,Gracey:2019xom,Kern:2019nzx,Aguilar:2019kxz,Kondo:2019rpa,Huber:2020keu,Nous:2020vdq,Aguilar:2019uob}. 
All approaches reveal a Landau gauge gluon propagator that is finite for all momenta including its zero momentum value.
This  means that QCD has a dynamical mechanism that generates mass scales and regularize the gluon propagator making it finite,
in contrast to the prediction of perturbation theory where it diverges at zero momentum. From the good agreement of all the non-perturbative 
results one can claim to have a good understanding of the gluon propagator from the low momentum region up to the ultraviolet limit.

For lattice simulations, the continuum limit of the gauge theory should be performed  to produce a proper estimation of the propagator 
\cite{Montvay:1994cy,Gattringer:2010zz}. However, the lattice computation of non-gauge invariant correlation functions requires rotating the links, 
obtained by importance sampling using e.g. the Wilson action, to a given gauge \cite{Giusti:2001xf}. This is a  very time consuming operation
from the computational point of view \cite{Oliveira:2003wa}.  The difficulties with the continuum extrapolation of the propagator comes not only from the 
gauge fixing process itself but also from the extrapolation towards the continuum limit of the propagators, which is a non trivial task \cite{Oliveira:2012eh}. 
Indeed, a change in the volume or in the number of lattice points changes the momentum accessed in the simulation, preventing a straightforward 
extrapolation towards the continuum limit. However, it is crucial to have good control on the systematics, i.e. the finite volume and lattice spacing effects referred
as lattice artefacts, to deliver a reliable propagator. On a typical measurement of a propagator, instead of performing the various limits, various lattice spacings 
and volumes are considered and the results compared to check for finite volume and lattice spacing effects.

The standard approach to the lattice evaluation of the Landau gauge gluon propagator assumes that the simulations are performed close to continuum physics
and, therefore, the tensor structure for the lattice propagator is that of the continuum theory. In momentum space, the propagator is written as
\begin{equation}
   D^{ab}_{\mu\nu} (p) = \delta^{ab} \, D_{\mu\nu} (p) = \delta^{ab} \, \left( \delta_{\mu\nu} - \frac{p_\mu p_\nu}{p^2} \right) \, D(p^2) \ ,
   \label{Eq:ContProp}
\end{equation}
where latin letters refer to colour indices and greek letters refer to Lorentz components.
It is well known that, due to the breaking of rotational invariance, the tensor structure of the lattice gluon propagator does not match the tensor
structure of the continuum theory given in Eq. (\ref{Eq:ContProp}). Indeed, the lattice data for the gluon dressing function 
$d(p^2) = p^2 \, D(p^2)$ reveals a structure \cite{Leinweber:1998uu}
that can be understood in  terms of the breaking of the rotational group $O(4)$ into $H(4)$ \cite{Becirevic:1999uc,deSoto:2007ht}, the 
symmetry group associated with an hypercubic lattice\footnote{This group includes rotations of multiples of $\pi/2$
around any  lattice axis and the corresponding reflection operations. More  about the $H(4)$ group later.}.

For the Landau gauge gluon propagator, two main approaches have been devised to handle the lattice artefacts  
in the results obtained assuming the continuum tensor structure (\ref{Eq:ContProp}). 
A common choice is to use an alternative definition of the lattice momentum combined with
a proper choice of the available momentum configurations, where only a subset of the kinematical configurations,
the momenta that are closer to the diagonal configuration $(p, \, p, \, p, \, p)$,  are considered. 
In the literature this procedure is named as momentum cuts  \cite{Leinweber:1998uu}. 
To have a better access to the deep infrared region, sometimes all  momenta  below a given threshold are also 
considered \cite{Oliveira:2012eh,Dudal:2018cli}. 
The introduction of  momentum cuts implies that a great deal of information on the propagator is lost. 

In what concerns the choice of momenta to reduce the effects due to the breaking of rotational symmetry, instead of using the lattice momenta, 
also named naive momentum below, 
\begin{equation}
   p_\mu = \frac{2 \, \pi}{a \, L_\mu} \, n_\mu \ , \qquad\qquad n_\mu = -\frac{L_\mu}{2} +1 , \, \dots, \, -1,  \, 0, \, 1, \, \dots , \,  \frac{L_\mu}{2}
   \label{Eq:Lat_mom}
\end{equation}
where $L_\mu$ is the number of lattice points along direction $\mu$ and $a$ is the lattice spacing, it is common to consider the improved lattice momenta
\begin{equation}
   \hat{p}_\mu = \frac{2 }{a} \, \sin\left( \frac{\pi}{L_\mu} \, n_\mu \right) \ , \qquad\qquad n_\mu = -\frac{L_\mu}{2} +1 , \, \dots, \, -1,  \, 0, \, 1, \, \dots , \, \frac{L_\mu}{2}
   \label{Eq:Lat_Imp_mom}
\end{equation}
that appears in the perturbative solution for the gluon propagator when the lattice is used as a regulator. 
Other definitions for the momentum have also been considered by several authors.
The use of the improved momentum combined with the momentum cuts reduces considerably the observed structures in $d(p^2)$ seen in  lattice simulations. 
Last but not least, in general 
the Landau gauge condition $p_\mu A_\mu (p) = 0$ is better fulfilled for the improved momentum, 
with $\left| \hat{p}_\mu A_\mu (p) \right| \ll \left| p_\mu A_\mu (p) \right|$ by several orders of magnitude. 

Another way to handle the lattice artefacts uses the lattice momentum $p_\mu$ and explores the invariants of the remnant $H(4)$ symmetry group
\cite{Becirevic:1999uc,deSoto:2007ht} associated with a hypercubic lattice
\begin{equation}
  p^2 = p^{[2]} = \sum_\mu p^2_\mu ,  \qquad
  p^{[4]} = \sum_\mu p^4_\mu ,  \qquad
  p^{[6]} = \sum_\mu p^6_\mu ,  \qquad
  p^{[8]} = \sum_\mu p^8_\mu \ .
  \label{Eq:ScalarInvarianst}
\end{equation}
Any other $H(4)$ invariant can be expressed in terms of $p^{[2]}$, $p^{[4]}$, $p^{[6]}$ and $p^{[8]}$ and, in this sense, the above invariants define 
the minimal set of lattice scalars in four dimensions. It follows that a lattice calculation of any scalar quantity $F$ is a function of all $H(4)$ invariants, i.e. 
$F_{Lat} = F(p^{[2]}, p^{[4]}, p^{[6]}, p^{[8]})$, and its continuum limit is given by $F(p^{[2]}, 0, 0, 0)$, module possible $p^2$ corrections.
If the lattice corrections are sufficiently small, extrapolations of $F_{Lat}$ to the continuum limit can be performed assuming that
$F_{Lat}$ can be written as a power series of the scalar invariants (\ref{Eq:ScalarInvarianst}).
This approach was applied successfully to the Landau gauge gluon propagator \cite{Becirevic:1999uc,deSoto:2007ht} 
and to the quark propagator \cite{Oliveira:2018lln}.
The extrapolation can not be applied to all  momenta accessed in a simulation as it requires data with the same 
$p^2$ but different $p^{[4]}$, $p^{[6]}$, $p^{[8]}$.
The infrared momenta and the highest momenta accessed in a simulation have a unique momenta for each $p^2$ and, therefore, for these momenta
the extrapolation can not be applied.

The procedures sketched above are the two main approaches to handle  lattice artefacts for the gluon propagator. However, there are other possibilites and,
for example, it is also possible to rely on lattice perturbation theory \cite{Capitani:2002mp,DiRenzo:2010cs} to estimate the corrections to the
non-perturbative lattice propagators and vertices \cite{Skullerud:2000un,Skullerud:2001aw,Skullerud:2003qu,Kizilersu:2006et,Oliveira:2018lln}. 

In the continuum formulation of QCD, the gluon propagator is a second order symmetric tensor, in Lorentz space, with respect to  the
transformations of the $O(4)$ group. 
The symmetry group of the lattice formulation of QCD is $H(4)$ and the lattice  gluon propagator is  a second order symmetric  tensor with respect to
the transformations associated with this group. Then, the
tensor structure of the lattice gluon propagator differs from
the continuum tensor structure given in Eq. (\ref{Eq:ContProp}).
For simulations close to the continuum limit, as those performed in the perturbative scaling regime, one expects the deviations from the continuum to be
small but they do not necessarily vanish.
In order to build a second order symmetric tensor that can be associated with the lattice gluon propagator 
it is  necessary to identify the lattice vectors, i.e. those quantities that behave as vectors with respect to $H(4)$ transformations, and then build the 
possible two dimensional symmetric tensors.

In color space, the lattice and the continuum propagators are 
second order  color tensors and the identity 
 $D^{ab}_{\mu\nu} (p) = \delta^{ab} D_{\mu\nu} (p)$ 
 holds for both formulations of QCD.
Indeed, $\delta^{ab}$ is the only symmetric second order SU(3) color tensor available.
It remains to identify the tensor basis that describes $D_{\mu\nu} (p)$,
a second order symmetric $H(4)$ tensor.
We aim to explore 
the tensor representations of $H(4)$ to measure 
the Landau gauge lattice gluon propagator in 4D simulations, to quantify
the lattice artefacts  and to see if
the tensor representations can improve the description of the lattice Landau gauge gluon propagator. 
As will be discussed below the use of the tensor representations also allows to test the Slavnov-Taylor identity for the gluon propagator 
with lattice simulations.

In the continuum formulation of QCD, the Slavnov-Taylor identity for the gluon determines its tensor structure. 
For the Landau gauge, this identity requires the gluon propagator to be orthogonal and described by a unique form factor as given by (\ref{Eq:ContProp}). 
In the lattice formulation of QCD, the continuum Slavnov-Taylor identity for the gluon does not necessarily apply. The orthogonality of the lattice Landau gluon propagator 
follows from the definition of the Landau gauge. However, the gauge definition does not imply a tensor structure for the gluon propagator as that 
given by (\ref{Eq:ContProp}). Indeed, as will be discussed below, the lattice Landau gluon propagator has more than a single form factor
and the definition of the Landau gauge, i.e. its orthogonality, implies that the form factors associated with the gluon propagator are not all independent. 
The relations between the form factors can be used to test for lattice  artefacts and also to check if the Slavnov-Taylor identity for the gluon is verified,
implying a unique form factor to describe the lattice gluon propagator.

The $H(4)$ tensor representations were introduced in \cite{Vujinovic:2019kqm} and used to explore and quantify the lattice artefacts effects in momentum 
space for two and three point correlation functions in pure Yang-Mills theories at lower space-time dimensions.
In \cite{Vujinovic:2018nqc} similar reasonings were applied to the lattice three gluon vertex for space-time dimensions lower than four.
These studies conclude that the class of momenta with the smaller lattice artefacts are those momenta configuration that are close to the diagonal 
configuration $(p, \, p, \, p, \, p)$. Their results confirms that the momentum cuts introduced in \cite{Leinweber:1998uu} provide a $D(p^2)$ with the smallest
lattice artefacts. A similar conclusion was also achieved in \cite{Becirevic:1999uc,deSoto:2007ht} when 
trying to understand the lattice artefacts based on $H(4)$ invariants.

In the current article, we explore the use of $H(4)$ tensor bases in the description of the lattice Landau gauge gluon propagator. Although 
the focus is on the Landau gauge, the procedure worked out can be extended to other gauges and other correlation functions. 
By using different tensor bases we test their faithfulness, i.e. how accurately they reproduce the lattice data, how many form factors are 
necessary to describe the lattice $D_{\mu\nu} (p)$ and test the orthogonality condition of the lattice data for the Landau gauge. 
By looking at the faithfulness of the tensor bases one also touches the problem of the evaluation of the lattice artefacts for the Landau gauge gluon propagator.
Furthermore, by combining the $H(4)$ tensor representations with extrapolations towards the continuum that explore the $H(4)$ scalar invariants mentioned above,
we have another look to the evaluation of the lattice artefacts.  The various approaches explored to describe the gluon propagator produce essentially the same 
results and are in good agreement with the standard analysis that relies on the use of momentum cuts. 
The combination of the various techniques also allow for an estimation of the theoretical uncertainty in the calculation. 
This theoretical uncertainty should be evaluated in a precision era scenario that we have been arriving in the computation of the lattice propagators. 
Moreover, as mentioned previously by calling for the use of $H(4)$ tensor representations, different properties associated with the definition 
of Landau gauge on the lattice can be tested, as the orthogonality of the gluon field  and how well the continuum 
Slavnov-Taylor identity for the gluon is fulfilled by the lattice data. 
In general, we find that the expected properties for the Landau gauge gluon propagator are well reproduced by the lattice data.

The manuscript is organised as follows. In Sec. \ref{Sec:1} we review the computation of the Landau gauge gluon propagator and introduce the definitions
to be used later on.  In Sec. \ref{Sec:2} the usual procedure to measure the lattice gluon propagator is described, together with the way the lattice data is 
treated before accessing the lattice form factors. The tensor bases used to investigate the lattice Landau gauge gluon propagator are mentioned, followed by a discussion on how to compute the form factors for the largest tensorial basis. The orthogonality condition of the Landau 
gauge gluon propagator for our largest tensor basis is also studied.
The results for the various tensor basis form factors, together with the test of the continuum inspired relations,
can be found in Sec. \ref{Sec:GPropTensor}, that also includes a comparison with the already published data for $D(p^2)$.
In Sec. \ref{Sec:Orho} the orthogonality of the propagator is discussed and related to the completeness of the tensor basis.
Finally, in Sec. \ref{Sec:ultima} we summarise our results and conclude.

\section{Getting the lattice gluon propagator \label{Sec:1}}

In the formulation of QCD on a spacetime lattice, the fundamental bosonic variables are the gauge links $U_\mu (x)$.
These are related to the gluon field $A_\mu (x)$ by
\begin{equation}
   U_\mu(x) = \exp \bigg\{ i \, g\, a \, A_\mu ( x + a \, \hat{e}_\mu / 2 ) \bigg\} \ ,
\end{equation}   
where $g$ is the bare coupling constant, $a$ is the lattice spacing and $\hat{e}_\mu$ the unit vector along direction $\mu$.
In order to compute the lattice Landau gauge propagators, after the importance sampling\footnote{For the lattice data analysed in the current work, the importance
  sampling was done with the Wilson gauge action.}
the gauge configurations are
 rotated to maximise, over the gauge orbits, the functional
\begin{equation}
    F[g; \, U]  = \frac{1}{N\, D \, V} \sum_{\mu, \mu} \Re \, \mbox{Tr} \bigg[ \, g(x) \, U_\mu (x) \, g^\dagger (x + a \, \hat{e}_\mu ) \bigg] \ ,
\end{equation}    
where $N$ is the number of colors, $D$ is the number of space-time dimensions, $V$ the number of lattice points and $g \in SU(3)$ are the matrices that
define a gauge transformation. In our work, for the maximisation of $F[g; U]$ we used a Fourier accelerated steepest descent  method
and the maximisation process was stopped when the average value, over the lattice, of the lattice equivalent to $| \partial A (x) | ^2$ is below $10^{-15}$. 
As will be discussed later, for this precision, the orthogonality of the gluon field (see the definition below)
and, therefore, of the gluon propagator, at least in a given range of momenta, seems to be enough to deliver information on the continuum propagator.

From the rotated links, the gluon field is computed using the definition
\begin{equation}
  A_\mu(x + a \, \hat{e}_\mu /2 ) = \left. \frac{ U_\mu (x) - U^\dagger_\mu (x)}{2 \, i \, a \, g} \right|_{\text{traceless}} \ ,
\end{equation}
 the momentum space gluon field is
\begin{equation}
  A_\mu( p) = \frac{1}{V} \sum_x e^{ -i \, p \cdot(x + a \, \hat{e}_\mu /2 )} \,  A_\mu(x + a \, \hat{e}_\mu /2 )  \ .
\end{equation}
and the lattice gluon propagator is given by
\begin{equation}
   \langle \, A^a_\mu (p) ~  A^b_\nu (p^\prime) \rangle = D^{ab}_{\mu\nu} (p) \, V \, \delta ( p + p^\prime)  \ ,
   \label{Eq:DefProp}
\end{equation}   
where $\langle \cdots \rangle$ stands for the ensemble average and
\begin{equation}
D^{ab}_{\mu\nu} (p)  = \delta^{ab} \, D_{\mu\nu} (p)
\end{equation}
where $p$ stands either for the naive lattice momentum (\ref{Eq:Lat_mom}) or the lattice improved momentum (\ref{Eq:Lat_Imp_mom}).
Further details on the definitions and on the gauge fixing procedure can be found in \cite{Silva:2004bv}.

The problem of the spacetime tensor structure of the gluon propagator is now reduced to the evaluation of the tensor decomposition
of $ D_{\mu\nu} (p)$. As already mentioned, on the lattice there is a minimum set of scalar invariants associated with the 
$H(4)$ symmetry group and the lattice gluon propagator can be written as
\begin{equation}
   D_{\mu\nu} (p) = \sum_i \, D^{(i)} (p^{[2]}, \, p^{[4]}, \, p^{[6]}, \, p^{[8]}) ~ \Ld^{(i)}_{\mu\nu}(p) \ ,
\end{equation}
where $D^{(i)} (p^{[2]}, \, p^{[4]}, \, p^{[6]}, \, p^{[8]})$ are $H(4)$ scalar form factors
and $\Ld^{(i)}_{\mu\nu}(p)$ are the elements of a tensor basis of operators built from the $H(4)$ tensors. 
It is only after the definition of the tensor basis that it is possible to access the form factors $D^{(i)}$, 
identify those $D^{(i)}$ that do not vanish in the continuum limit and quantify their deviations from the continuum limit due to the lattice artefacts.

\section{Tensor bases for the gluon propagator \label{Sec:2}}

The elements of the hypercubic symmetry group $H(4)$ are rotations by $\pi / 2$, rotations by multiples of $\pi / 2$ around any of the hypercube axes 
and the corresponding reflection operations.
As discussed in \cite{Vujinovic:2019kqm} the vectors with respect to $H(4)$ transformations
are the naive lattice momentum $p_\mu$, the improved lattice momentum $\hat{p}_\mu$, 
or any odd power of these quantities. Although in this section we will use the notation $p_\mu$ to refer to a vector, the reader should keep in mind that $p_\mu$ 
can be read either the naive momentum $p$,  the improved lattice momentum $\hat{p}$, or any odd power of any of these momenta. 
In the following discussion, we ignore the case where products of different types of vectors are considered. This is not a limitation of the approach
as the improved momenta can be written as a power series of odd powers of the naive momenta.

In our discussion of the Landau gauge gluon propagator we will consider different tensor bases, including the continuum tensor basis (\ref{Eq:ContProp}).
The reported data will be compared with the single form factor $D(p^2)$ associated with the continuum basis (\ref{Eq:ContProp}) obtained
applying the momentum cuts, where $D(p^2)$ is described as a function of the improved momenta (\ref{Eq:Lat_Imp_mom}).
This data that will be used here as reference data for the propagator was published in \cite{Dudal:2018cli}. 
Note that in this work the propagator was calculated performing a 
$Z_4$ average where, for each gauge configuration, an 
average over equivalent momenta, obtained from permutations over the momentum components, was performed.

Herein, we recompute the propagator relying on the continuum tensor basis and exploring the $H(4)$ invariants. In the new calculation
the propagator was computed as function of $p$ and evaluated after (i) grouping all the data points with the same set of scalar invariants, 
(ii) for each gauge configuration perform a data average on this equivalent class of data and only then compute the form factors and, finally, 
(iii) do the ensemble average. This approach, which explores the $H(4)$ symmetry group, produces a much clearer and smoother signal for the 
propagator due to the increase on the number of operations that wash out the fluctuations of the Monte Carlo simulation.
Furthermore, for the case where the analysis of the lattice data is performed in terms of the lattice momentum $p$, a linear extrapolation in $p^{[4]}$, 
towards $p^{[4]} = 0$, for the form factors will also be considered. We recall the reader that the extrapolation is not possible for the lowest and for the 
highest momenta. The extrapolation ignores the dependence on $p^{[6]}$ and on $p^{[8]}$ of the form factors and will also take the various $p^{[4]}$ 
data points as independent variables. 
The values reported for the form factors after the extrapolation are the predictions of a linear regression at $p^{[4]} = 0$, 
with the quoted errors being those of the linear regression, i.e. the statistical errors for the final results do not rely on bootstrapping.
This difference in the estimation of the statistical errors underestimates the fluctuations and therefore the reported errors for the final results are smaller.

The standard approach to compute the Landau gauge lattice gluon propagator assumes that the continuum tensor structure (\ref{Eq:ContProp}) also
describes the lattice propagator. Then,  the unique form factor can be computed 
from\footnote{Note that a different normalization constant is used for zero momentum.}
\begin{equation}
   D(p^2) = \frac{1}{(N^2-1) \, (D-1)} \sum_{a, \, \mu} D^{aa}_{\mu\mu}(p) \ .
      \label{Eq:LandauD}
\end{equation}
The above relation (\ref{Eq:LandauD}) holds independently of the choice of momenta being $p$ or $\hat{p}$. 
The lattice gluon propagator is a function not only of $p^2$ but of all $H(4)$ scalars. However, we have omitted this dependence in Eq. (\ref{Eq:LandauD})
to simplify the notation.

Following \cite{Vujinovic:2019kqm},  a minimal tensor basis to describe the lattice propagator reads
\begin{eqnarray}
D^{ab}_{\mu\mu} (p) & =  & \delta^{ab} \bigg( E(p^2) \, \delta_{\mu\mu} + F(p^2) \, {p}^2_\mu  \bigg) ~  \ , \qquad \mbox{ (no sum) } 
\nonumber \\
D^{ab}_{\mu\nu} (p) & =  & \delta^{ab} ~  H(p^2) \, {p}_\mu {p}_\nu   ~ \ , \qquad \mu\ne\nu \ .
   \label{Eq:LandauLattProp0}
\end{eqnarray}
All the three form factors are functions of the full set of lattice scalars $p^{[n]}$
with $n = 2, \, 4, \, 6, \, 8$ 
but, to simplify the notation, only the $p^{[2]} = p^2$ dependence of the form factors is written explicitly. 
This minimal tensor basis for the gluon propagator is independent of the  gauge. 
The gauge condition establishes a relation between the form factors and, therefore, the total number of independent form factors for this basis is smaller than three. 
For the Landau gauge we expect to recover $F(p^2) = H(p^2) = - E(p^2)/ p^2$ to reproduce the continuum tensor structure that is seen in
Eq. (\ref{Eq:ContProp}).

The $H(4)$ tensor bases can be further extended to include higher powers of the momenta. The simplest extension of (\ref{Eq:LandauLattProp0}), 
already mentioned in \cite{Vujinovic:2019kqm}, being
\begin{eqnarray}
D^{ab}_{\mu\mu} (p) & =  & \delta^{ab} \bigg( E(p^2) \, \delta_{\mu\mu} + F(p^2) \, {p}^2_\mu +  G(p^2) \,  {p}^4_\mu \bigg) ~  \ , \qquad \mbox{ (no sum) } 
\nonumber \\
D^{ab}_{\mu\nu} (p) & =  & \delta^{ab} ~  \bigg( H(p^2) \,  {p}_\mu \, {p}_\nu + I(p^2) \,  {p}_\mu \, {p}_\nu  \left(   {p}^2_\mu +  {p}^2_\nu \right) \bigg) ~ \qquad \mu\ne\nu 
   \label{Eq:LandauLattProp}
\end{eqnarray}
and requires five form factors.  
A straightforward algebra shows that the projectors to extract the form factors that are associated with the diagonal components of the propagator read
\begin{small}
\begin{equation}
E(p^2)  =    \frac{
    \left( \sum_\mu D_{\mu\mu} (p)  \right) \left( p^{[4]}  p^{[8]} - \left(  p^{[6]} \right)^2   \right) 
+  \left( \sum_\mu p^2_\mu D_{\mu\mu}(p) \right) \left(   p^{[4]}  p^{[6]} - p^2  p^{[8]} \right)  
+  \left( \sum_\mu p^4_\mu D_{\mu\mu}(p) \right) \left(   p^2 p^{[6]} - \left( p^{[4]}  \right)^2  \right)  
}{ \Delta_1 } \ ,
\label{Eq:Proj_E}
\end{equation}
\begin{equation}
F(p^2)  =    \frac{
    \left(\sum_\mu D_{\mu\mu} (p)\right)  \left( p^{[4]}  p^{[6]} - p^2  p^{[8]}  \right) 
+  \left( \sum_\mu p^2_\mu D_{\mu\mu}(p) \right) \left(   d \,   p^{[8]} -  \left( p^{[4]} \right)^2 \right)  
+  \left( \sum_\mu p^4_\mu D_{\mu\mu}(p) \right) \left(   p^2  p^{[4]} -  d \, p^{[6]} \right)  
}{ \Delta_1 } 
   \label{Eq:Proj_F}
\end{equation}
\end{small}
and
\begin{small}
\begin{equation}
G(p^2)  =    \frac{
   \left(  \sum_\mu D_{\mu\mu} (p) \right)  \left( p^2  p^{[6]} - \left(  p^{[4]} \right)^2   \right) 
+  \left( \sum_\mu p^2_\mu D_{\mu\mu}(p) \right) \left(   p^2  p^{[4]} - d \, p^{[6]} \right)  
+ \left(  \sum_\mu p^4_\mu D_{\mu\mu}(p) \right) \left(   d \,   p^{[4]} - \left( p^2  \right)^2  \right)  
}{ \Delta_1 } \ ,
\end{equation}
\end{small}
where
\begin{equation}
 \Delta_1 = d \, \left( p^{[4]}  p^{[8]} - \left(  p^{[6]} \right)^2  \right) ~
+  ~ p^2 \, \left(   p^{[4]}  p^{[6]} - p^2  p^{[8]} \right)   ~ 
+  ~ p^{[4]} \,  \left(   p^2  p^{[6]} - \left( p^{[4]}  \right)^2  \right)  \ .
\label{Eq;delta_1}
\end{equation}
The projectors to compute the form factors for the components with $\mu \ne \nu$ are
\begin{equation}
H(p^2)  =    \frac{
    2 \, \left( \sum_{\underset{\mu\ne\nu}{\mu,\nu}} p_\mu p_\nu D_{\mu\nu} (p)  \right) \left( p^{[4]}  p^{[6]} -  p^{[10]}  \right) 
- 2 \,  \left( \sum_{\underset{\mu\ne\nu}{\mu,\nu}} p^3_\mu p^3_\nu D_{\mu\nu}(p) \right) \left(   p^2  p^{[4]} - p^{[6]} \right)  
}{ \Delta_2 } \ ,
\label{Eq:Proj_H}
\end{equation}
\begin{equation}
I(p^2)  =    \frac{
    \left(\sum_{\underset{\mu\ne\nu}{\mu,\nu}} p_\mu p_\nu D_{\mu\nu} (p)\right)  \left( p^{[8]} - \left( p^{[4]} \right)^2  \right) 
+  \left( \sum_{\underset{\mu\ne\nu}{\mu,\nu}} p^3_\mu p^3_\nu  D_{\mu\nu}(p) \right) \left(    \left( p^2 \right)^2  - p^{[4]} \right)  
}{ \Delta_2 } \ ,
   \label{Eq:Proj_I}
\end{equation}
with
\begin{equation}
  \Delta_2 = 2 \, \left( p^2 p^{[4]}  - p^{[6]} \right) \left( p^{[8]} - \left( p^{[4]} \right)^2 \right) ~ + ~2 \, 
  \left( \left( p^2 \right)^2 - p^{[4]} \right) \left( p^{[4]}  p^{[6]} -  p^{[10]} \right)  \ .
  \label{Eq;delta_1}
\end{equation}
The full set of form factors cannot be computed for all kinematics. For example, it follows
from the definitions of $\Delta_1$ and $\Delta_2$ that for on-axis and for diagonal momenta $\Delta_1 = \Delta_2 = 0$ 
and, therefore, the above relations cannot be used for these two classes of kinematic configurations. 
Moreover, for momenta of type $(m, \, n, \, n, \, n)$ and  $(m, \, m, \, n, \, n)$, in four dimensions, $\Delta_1 = 0$ and it is not possible to 
access independently $E(p^2)$, $F(p^2)$ and $G(p^2)$. 

The form factor $D(p^2)$ that describes the gluon propagator when  one uses the continuum tensor basis is a combination of $E(p^2)$ , $\dots$, $G(p^2)$.
From its definition, it comes that
\begin{eqnarray}
   D(p^2) & = &  \frac{1}{3} \bigg(  4 \, E(p^2, p^{[4]},  p^{[6]},  p^{[8]} ) +
                          F(p^2, p^{[4]},  p^{[6]},  p^{[8]} ) \, p^2 + G(p^2, p^{[4]},  p^{[6]},  p^{[8]} ) \, p^{[4]} \bigg) \\
              & \approx & \frac{1}{3}  \Bigg( 4 \, E(p^2, 0,  0,  0 ) +  F(p^2, 0,  0,  0) \, p^2  \nonumber \\
                    & & \qquad\qquad 
                            +  ~ \left(  4 \,  \frac{ \partial E(p^2, 0,  0,  0)}{\partial  p^{[4]}} + \frac{ \partial F(p^2, 0,  0,  0)}{\partial  p^{[4]}}  \, p^2 + G(p^2, 0,  0,  0 ) \right) \, p^{[4]} 
                            + \dots \Bigg) \nonumber \\           
   & = & E(p^2) + G(p^2) \, \frac{p^{[4]}}{3}  \ ,
   \label{Eq:Cont_D_ExtBas_Real}
\end{eqnarray}
where in the last line the continuum relation $F(p^2) \, p^{[2]} = - E(p^2)$  was used, and a weak dependence on $p^{[4]}$ for $F$ and $E$ was assumed.
If $G(p^2)$ is small enough this relation tell us that $E \approx D$.
We will check the validity of this relation in Sec. \ref{Sec:GPropTensor}.

\subsection{Orthogonality constraints for a general kinematical configuration \label{Sec:orto_gen}}

In the Landau gauge the gluon field is orthogonal to its momentum, i.e. $p \cdot A(p) = 0$. This condition constraints
the tensor structure of the propagator.
For the tensor basis considered in Eq. (\ref{Eq:LandauLattProp}), the orthogonality condition requires that
\begin{eqnarray}
   \sum_\mu p_\mu D_{\mu\nu} (p) & = & E(p^2)  \, p_\nu + F(p^2) \, p^3_\nu + G(p^2) \, p^5_\nu \nonumber \\
    & & \qquad + ~ H(p^2) \left( p^2 - p^2_\nu \right) \, p_\nu + I(p^2) \left( \left( p^{[4]} - p^4_\nu \right) p_\nu + \left(p^2 - p^2_\nu\right) p^3_\nu \right) = 0 \ .
    \label{Eq:TransvLatt}
\end{eqnarray}
If $p_\nu = 0$, this condition is automatically satisfied. However, if $p_\nu \ne 0$ then Eq. (\ref{Eq:TransvLatt}) translates into
\begin{eqnarray}
    E(p^2)  ~ + ~ F(p^2) \, p^2_\nu ~ + ~ G(p^2) \, p^4_\nu  ~ + ~  H(p^2) \left( p^2 - p^2_\nu \right)  ~ + ~ I(p^2) \left(  p^{[4]} + p^2\, p^2_\nu   -  2 \, p^4_\nu   \right) 
     = 0  \ ,
    \label{Eq:TransvLatt2}
\end{eqnarray}
a relation between the various form factors that can be tested on a lattice simulation. The computation of (\ref{Eq:TransvLatt2}) using the
lattice form factors also tests the completeness of the tensor basis.

\subsection{Special Kinematical Configurations \label{Sec:SpecialKinematics}}

As mentioned before, there are special momentum configurations where it is not possible to access all the form factors  of the tensor basis given in
(\ref{Eq:LandauLattProp}).
For \underline{on-axis momenta $(p, \, 0, \, 0, \, 0)$} the propagator is diagonal and  its spacetime tensor structure reads
\begin{equation}
  \bigg( D_{\mu\mu} (p) \bigg) = \bigg( E(p^2) + F(p^2) p^2 + G(p^2) p^4, ~~ E(p^2),  ~~ E(p^2), ~~ E(p^2) \bigg)\qquad\qquad\mbox{ (no sum) } \ .
\end{equation}
Then, only the following combinations of the lattice form factors 
\begin{equation}
   E(p^2) = \sum^{4}_{\mu = 2} \frac{D_{\mu\mu} (p) }{3} \ ,  \qquad\quad\mbox{ and } \qquad\qquad
   F(p^2) \, p^2 + G(p^2) \, p^4 = D_{11} (p) - E(p^2)
   \label{Eq:RefOnAxis}
\end{equation}
can be computed in a simulation. In the continuum limit $E(p^2)$ and $F(p^2) \, p^2$ are identified with $D(p^2)$, and $G(p^2) \, p^4$ measures
the deviations of the lattice propagator from its ``continuum" tensorial structure. The form factor measured in (\ref{Eq:LandauD}) is
\begin{eqnarray}
   D(p^2) & = & \frac{4 \, E(p^2) + F(p^2) p^2 + G(p^2) p^4}{3} 
    =  E(p^2) + \frac{G(p^2) p^4}{3} 
    \label{Eq:Mais_uma_OnAxis}
\end{eqnarray}
where in the last term the continuum inspired relation $p^2 F(p^2) = - E(p^2)$ was used.

If the orthogonality condition $p \cdot A(p) = 0$  holds, it requires that
\begin{equation}
  \sum_\mu p_\mu D_{\mu\nu} (p) =  E(p^2)\,p + F(p^2) \, p^3 + G(p^2)  \, p^5 = 0 \ .
\end{equation}  
This expression allows to test the orthogonality condition associated with the Landau gauge for  on-axis momenta. 
If $G(p^2)$ is small enough, see Sec. \ref{Sec:GPropTensor},
the orthogonality condition implies $F(p^2) = - E(p^2)/p^2$ and the usual tensor structure of the continuum theory is  recovered.
Note that in this case $D(p^2) = E(p^2)$ with  a correction given by $G(p^2) p^4$.
For on-axis momenta, the non-diagonal components of the propagator vanish both for the continuum theory and for its lattice formulation 
as they are proportional to $p_\mu p_\nu$ with $\mu \ne \nu$. 
Unless $G(p^2) \, p^4$ diverges for small $p^2$, the deviations from the continuum tensor structure for small momenta should be small or even vanish for this class
of momenta. The data for the gluon propagator for two and three dimensional theories reported in \cite{Vujinovic:2019kqm} suggest that the 
lattice gluon propagator recovers the continuum tensor structure for on-axis momenta, i.e. the data supports a small finite value for 
$G(p^2)$.

Let us now turn our attention to
\underline{diagonal momenta $(p, \, p, \, p, \, p)$}.
 Recall that, in lattice simulations and having in mind the suppression of lattice artefacts, this the preferred class of momenta used to measure $D(p^2=p^{[2]})$.
The diagonal elements of the propagator are
\begin{equation}
  D_{\mu\mu} (p)   =    E(p^{[2]}) + F(p^{[2]}) \, p^2 + G(p^{[2]}) \, p^4   \qquad\qquad\mbox{ (no sum)} \ , 
  \label{Eq:RefDiagonalDiagonal}
\end{equation}
while the non-diagonal elements read
\begin{equation}
  D_{\mu\nu} (p)   =    p^2 \bigg( H(p^{[2]}) + 2 \, I(p^{[2]})  \, p^2 \bigg)   \qquad\qquad \mu \ne \nu \ .
  \label{Eq:RefDiagonalNonDiagonal}
\end{equation}
The combination of the form factors given in (\ref{Eq:RefDiagonalDiagonal}) and (\ref{Eq:RefDiagonalNonDiagonal}) are the quantities that can be measured in 
a lattice simulation. The continuum limit implies that $E(p^{[2]}) \rightarrow D(p^{[2]})$ and $4 \, p^2 \, F(p^{[2]}) \rightarrow  4 \, p^2 H(p^{[2]}) \rightarrow - D(p^{[2]})$
and that the contributions associated with $G$ and $I$ should vanish as $a \rightarrow 0$.

For diagonal momenta, the orthogonality condition relates the measurable form factors as
\begin{equation}
  \sum_\mu p_\mu \, D_{\mu\nu} (p)   =   p \bigg( E(p^{[2]}) + F(p^{[2]}) \, p^2 + G(p^{[2]}) \, p^4  + 3 \, p^2 \, H(p^{[2]}) + 6 \, I(p^{[2]})  \,  p^4    \bigg) = 0 
\end{equation}
or
\begin{equation}
    \bigg( E(p^{[2]})  + F(p^{[2]}) \, p^2 + G(p^{[2]}) \, p^4  \bigg) + 3 \, p^2 \, \bigg( H(p^{[2]}) \, + 2 \, I(p^{[2]}) \, p^2 \bigg)  = 0 \ .
    \label{Eq:OrtoLattDiag}
\end{equation}
Again, a measure of this quantity tests the Landau gauge condition on the lattice. If the contribution of $p^4 \, G(p^{[2]}) $ and of $ p^4 \,I(p^{[2]})$ is small or negligible, 
this condition becomes
\begin{equation}
    E(p^{[2]})  +  p^2 \, \bigg(  F(p^{[2]}) \, + 3 \,  H(p^{[2]})  \bigg)  = 0 \ .
    \label{Eq:OrtoLattDiag}
\end{equation}
Further, if $F(p^{[2]}) = H(p^{[2]})$ holds, the continuum tensor structure is recovered for the diagonal momenta. 
Unfortunately, for this kinematical configuration it is impossible to
access all the form factors and the only test that can be performed is the orthogonality condition as given in Eq. (\ref{Eq:OrtoLattDiag}).

The on-axis and the diagonal momenta are not the only kinematical configurations that have $\Delta_1 = 0$ or $\Delta_2 = 0$ and prevent the use of the
projectors (\ref{Eq:Proj_E}) - (\ref{Eq:Proj_I}) to measure all the form factors.  
For \underline{momenta of type $p = (a, \, b, \, b, \, b)$}, that have $p^2 = a^2 + 3\, b^2$,
 it turns out that, in four dimensions, $\Delta_1 = 0$ and $\Delta_2 = 36 \, a^2 b^8 (b^2 - a^2)^2$. 
The form factors $H(p^2)$ and $I(p^2)$ can be computed as usual, but not the remaining ones. Looking at the Lorentz components of the gluon propagator,
see Eq. (\ref{PropLorentzStruct}) in App. \ref{Prop:LatExtBasis}, it follows that
\begin{equation}
   D_{11}(p) = E(p^2) + F(p^2) \, a^2 + G(p^2) \, a^4  \quad\mbox{ and }\quad
   D_{22}(p) = D_{33}(p) = D_{44}(p) = E(p^2) + F(p^2) \, b^2 + G(p^2) \,b^4 
\end{equation}
are the only combinations related to the diagonal components that can be measured directly in a simulation.
The continuum tensor structure implies  $E(p^2) = - \, p^2 \, H(p^2)$. 
The form factor $H(p^2)$ can be measured using the off-diagonal components and, combined with the diagonal components, it is possible to access both
 $G(p^2)$ and $F(p^2)$. This last form factor should reproduce $H(p^2)$, if the lattice propagator reproduces the continuum tensor structure. 
 For this kinematical configuration, the orthogonality of the propagator translates into the two conditions
\begin{equation}
\left\{ 
\begin{array}{lrrl}
  \left[ ~ E(p^2)  + a^2 F(p^2) + a^4 G(p^2) ~ \right]  ~ + &   b^2 \, H(p^2)  ~+  & b^2  (a^2 + b^2) \, I(p^2)    &   = 0 \ , \\
  \\ 
  \left[ ~  E(p^2) + b^2 F(p^2) + b^4 G(p^2) ~ \right] ~+ &    \left( a^2  + 2 \, b^2 \right) \, H(p^2) ~+   &   \left( a^2  (a^2 + b^2)  + 4 \, b^4 \right) \, I(p^2)    &  = 0  \ \\
\end{array}
\right.
\end{equation}
that can be tested in lattice simulations.  
For \underline{momentum of type $p = (a, \, a, \, b, \, b)$}, that have $p^2 = 2 ( a^2 + b^2)$, the denominators in Eqs. (\ref{Eq:Proj_E}) and (\ref{Eq:Proj_H}) 
are $\Delta_1 = 0$ and  $\Delta_2 = 8 \, a^2 \, b^2 \, (b-a)^2 \, (b + a)^2 \, (b^2 + a^2) \, ( 2 \, b^4 - a^2 b^2 + 2 \, a^4)$, respectively.
As in the previous case, the projectors Eqs. (\ref{Eq:Proj_H}) and (\ref{Eq:Proj_I}) can be used to measure $H(p^2)$ and $I(p^2)$.
The diagonal terms of the propagator are given by
\begin{equation}
      D_{11}(p) = D_{22} (p) = E(p^2) + F(p^2) \, a^2 + G(p^2) \, a^4 \qquad\mbox{ and }\qquad
      D_{33}(p) = D_{44} (p) = E(p^2) + F(p^2) \, b^2 + G(p^2) \, b^4 
      \label{Eq:momaabb}
\end{equation}
and  these are the only combinations of form factors that can be measured in a lattice simulation for this class of momenta.
Continuum physics requires $E(p^2) = - \, p^2 H(p^2)$ and, assuming that this condition holds, $F(p^2)$ and $G(p^2)$ can be measured
using Eqs. (\ref{Eq:momaabb}). The continuum tensor structure of the propagator requires also that $F(p^2) = H(p^2)$ and, in this case,
the two quasi-independent measures of $F$ and $H$ can be used to test how well the continuum basis describes the lattice data. 
The orthogonality condition relates the form factors and these relations can be used to test, once more,
the outcome of the simulations.
The last special kinematical configuration that we have identified that belongs to the  \underline{class of kinematics under analysis is $p = (a, \, b, \, 0, \, 0)$} 
that  has $p^2 = a^2 + b^2$. The non-vanishing components of the gluon propagator are 
\begin{equation}
\begin{array}{l@{\hspace{0.5cm}}l}
    D_{11}(p) = E(p^2) + a^2 F(p^2) + a^4 G(p^2) \ ,  &    D_{22}(p) = E(p^2) + b^2 F(p^2) + b^4 G(p^2) \ ,  \\
    \\
    D_{33}(p) = D_{44}(p) = E(p^2)  \ ,  &     D_{12}(p) = a \, b \, \left[ H(p^2) + I(p^2) ( a^2 + b^2) \right]
    \end{array}
\end{equation}
and, therefore, it is only possible to measure $E(p^2)$, $F(p^2)$, $G(p^2)$ and $H(p^2) + I(p^2) ( a^2 + b^2)$ directly in a lattice simulation. 
If one assumes that the continuum relations hold and set $H(p^2) = F(p^2)$, then one can extract the form factor $I(p^2)$. 
Once more, the Landau gauge orthogonality condition translates into two equations for the full set of form factors that can be tested with lattice data.

All the relations that we have mentioned so far assume that the simulations have infinite statistics and, therefore, that they can be used to measure
all the form factors and/or its combinations. In a real simulation, where only a finite number of gauge configurations is accessed for each ensemble,
the limited statistics translates into large statistical errors for some of the form factors and, in practice, some of the relations derived
do not provide any valuable information on the lattice propagator.

\section{The Landau Gauge Lattice Gluon Propagator \label{Sec:GPropTensor}} 

As stated before, one of the goals of the current work is to check whether one can arrive at a faithful description of the lattice gluon propagator. 
To do so, we explore various tensor bases of operators and measure the associated form factors. We will investigate
the continuum tensor basis given in Eq. (\ref{Eq:ContProp}), i.e.
\begin{displaymath}
D^{ab}_{\mu\nu} (p) =  \delta^{ab} \, D_{\mu\nu} (p) = \delta^{ab} \, \left( \delta_{\mu\nu} - \frac{p_\mu p_\nu}{p^2} \right) \, D(p^2) \ ,
\end{displaymath}
that requires a unique form factor, the modified continuum basis
\begin{equation}
	D_{\mu\nu}(p) = A(p) \, \delta_{\mu\nu} + B(p) \, p_\mu p_\nu \ ,
	\label{eq:lattice_continuum_basis}
\end{equation}
that calls for two form factors, an extended basis
\begin{eqnarray}
& & D_{\mu\mu}(p) = J(p) \, \delta_{\mu\mu} + K(p) \, p_\mu^2, \qquad \mbox{ (no sum) } .\nonumber \\
&& D_{\mu\nu}(p) = L(p) \, p_\mu p_\nu, \qquad\qquad\qquad \mu\neq\nu  \ ,
\label{eq:partial_lattice_basis}
\end{eqnarray} 
that uses three form factors, and the enlarged tensor basis given in Eq. (\ref{Eq:LandauLattProp}), i.e.
\begin{eqnarray}
D_{\mu\mu} (p) & =  &  E(p^2) \, \delta_{\mu\mu} + F(p^2) \, {p}^2_\mu +  G(p^2) \,  {p}^4_\mu  ~  \ , \qquad \mbox{ (no sum) } 
\nonumber \\
D_{\mu\nu} (p) & =  &  H(p^2) \,  {p}_\mu \, {p}_\nu + I(p^2) \,  {p}_\mu \, {p}_\nu  \left(   {p}^2_\mu +  {p}^2_\nu \right) ~ \qquad \mu\ne\nu  \ ,
\nonumber 
\end{eqnarray}
that requires five different form factors. 
All form factors are function of all the lattice scalar invariants but to simplify the notation we only write explicitly the $p^2$ dependence.

The various tensor basis representations will be analysed using the lattice data for the Landau gauge gluon propagator
generated with the Wilson action, at $\beta = 6.0$, for a Monte Carlo simulation performed on a $80^4$ lattice and that uses 550 gauge configurations. 
For this simulation the lattice spacing, measured from the string tension \cite{Bali:1992ru}, is $a = 0.1016(25)$ fm or $1/a = 1.943(47)$ GeV. 
In physical units the length of each lattice side is $8.13(20)$ fm. In all cases but the $H(4)$ extrapolation statistical errors are computed with the bootstrap method
with a 67.5\% confidence level.
The quoted errors for the $p^{[4]}$ extrapolation are the exception and the numbers reported are those obtained in a linear regression.

\subsection{Results with the continuum tensor basis and diagonal like momenta \label{Sec:GetD}} 

\begin{figure}[t] 
   \centering
   \includegraphics[width=4in]{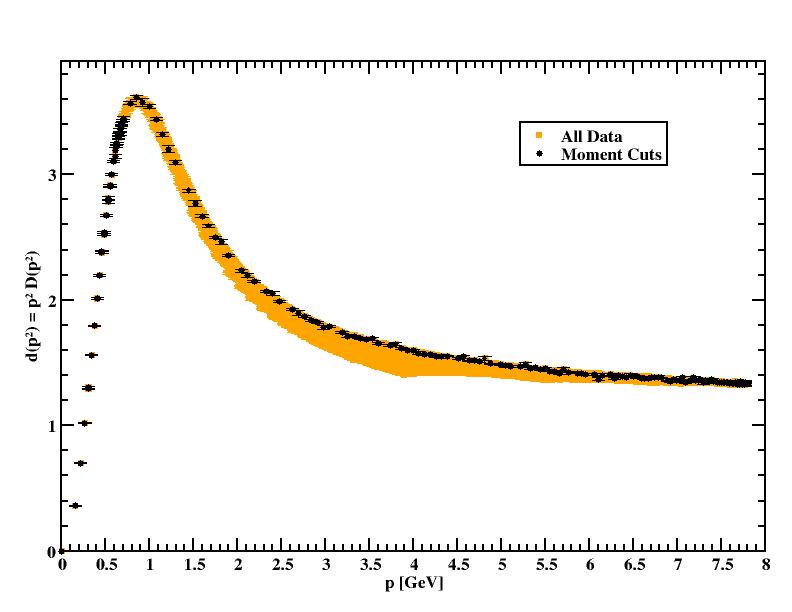} 
   \caption{The gluon dressing function $d(p^2) = p^2 D(p^2)$ as a function of the improved momentum for the continuum tensor basis and for
                 the momentum configurations that verify the momentum cuts and all the kinematical configurations.}
   \label{fig:FF_Cont_Basis_NoBin}
\end{figure}

Let us start our analysis looking at the results for the Landau gauge gluon propagator based on the use of continuum tensor  basis
(\ref{Eq:ContProp}). The propagator is described by a single form factor $D(p^2)$, whose value is computed with the help of Eq. (\ref{Eq:LandauD}). 
We will report different calculations for $D(p^2)$ together with the data already published in \cite{Dudal:2018cli}, that will be used as reference data. 
The evaluation of the reference data relies on the $Z_4$ average of equivalent momenta, with $D(p^2)$ given in terms of the improved lattice momenta $\hat{p}_\mu$ 
and only the subset of those lattice momenta that verify the cylindrical and conical cuts defined in \cite{Leinweber:1998uu}, for momenta above 0.7 GeV, is considered. 
For the infrared momenta, i.e for those momenta below 0.7 GeV, all available lattice data is taken into account.

In Fig. \ref{fig:FF_Cont_Basis_NoBin} we show the gluon dressing function for the data published in \cite{Dudal:2018cli}, i.e. those momentum
configurations that satisfy the cuts (black points), together with all the momenta accessed in the simulations (orange points).
The Fig. shows the spread of the data due to the lattice artefacts for the full range of improved lattice momenta and 
illustrates  the type of data that is selected by the momentum cuts.

\begin{figure*}[t] 
   \centering
   \includegraphics[width=3.3in]{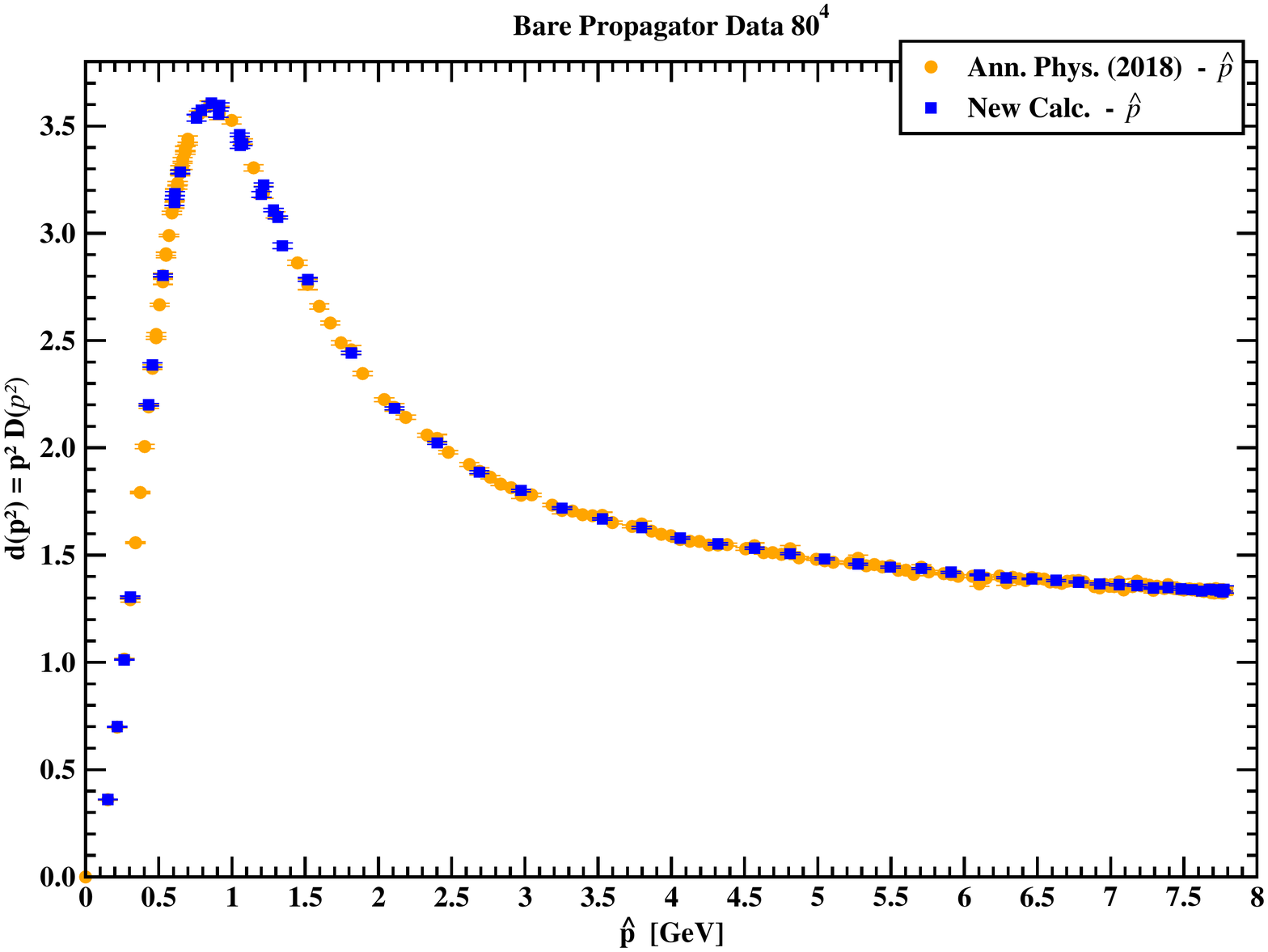}  
   \includegraphics[width=3.3in]{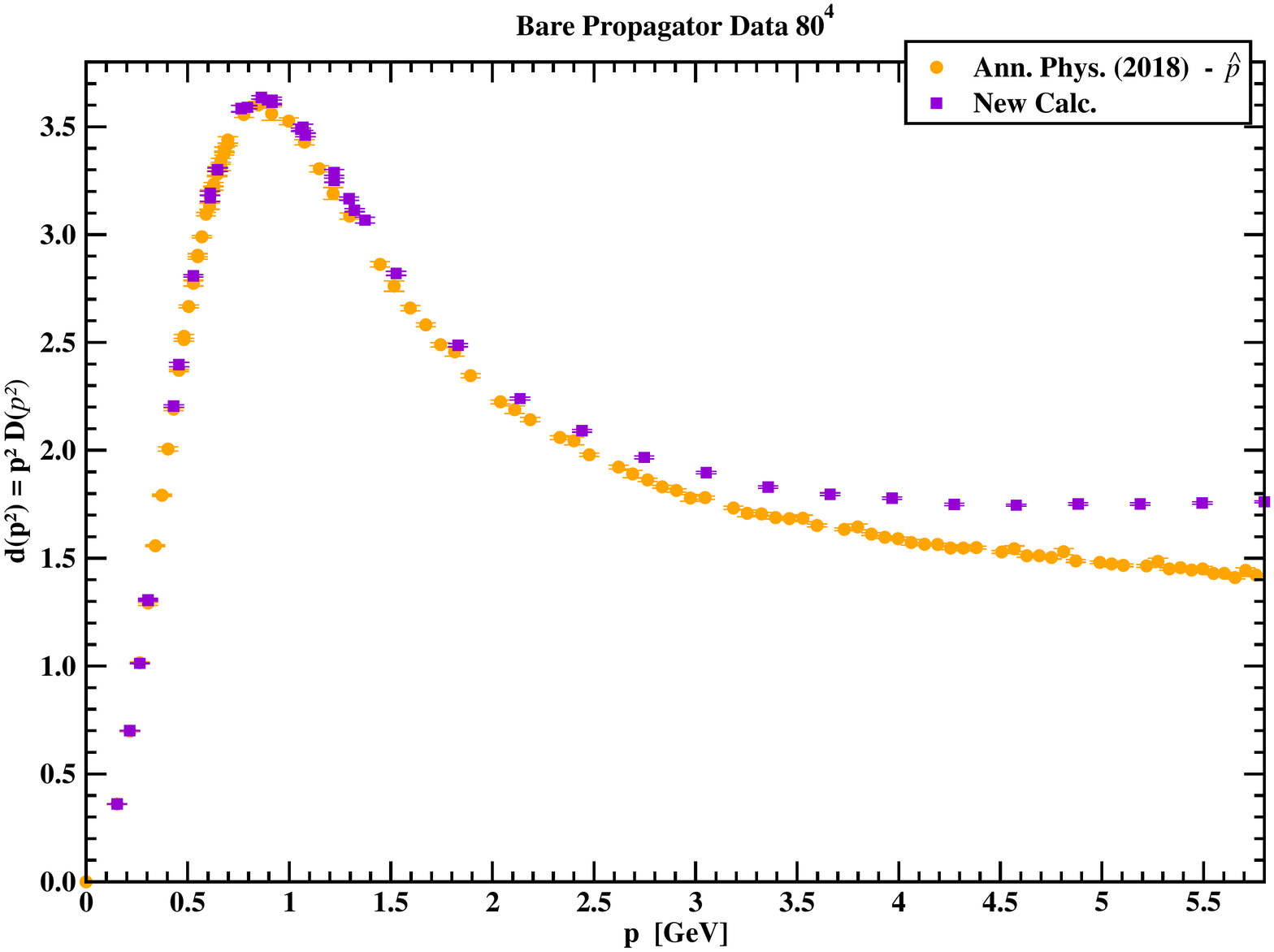}  \\
   \includegraphics[width=3.3in]{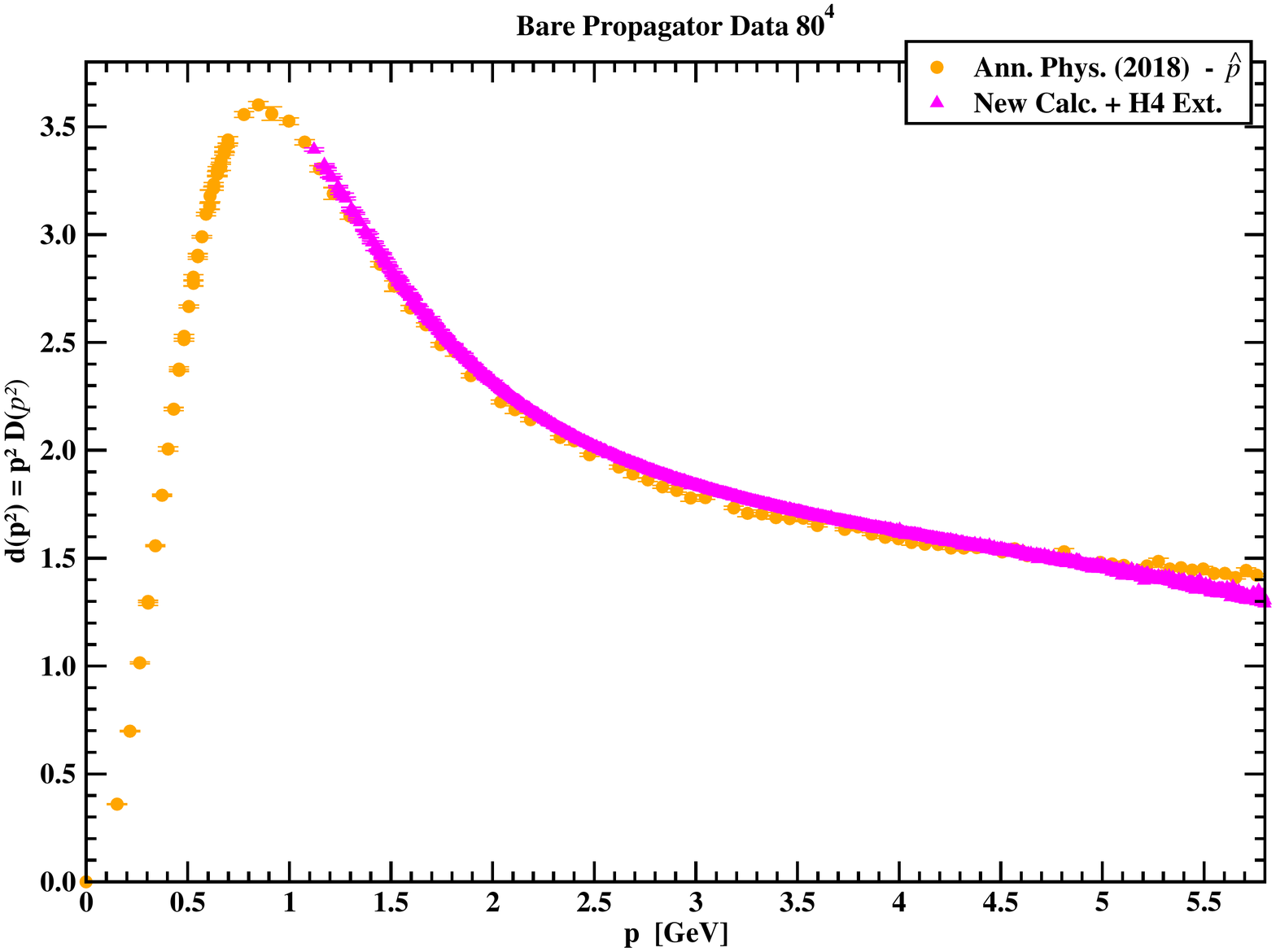}
   \caption{The Landau gauge gluon dressing function computed using the continuum like tensor basis written in terms of the improved momenta (top-left), in terms 
                  of the lattice momenta (top-right) and after performing the $H(4)$ extrapolation (bottom). Note that the data with the $H(4)$ extrapolation
                  covers a smaller range of $p$. See text for details.}
   \label{fig:LatContBas80}
\end{figure*}

The lattice gluon dressing function $d(p^2) = p^2 D(p^2)$ is shown in Fig. \ref{fig:LatContBas80} for different definitions of the momenta that appear in the projectors. 
The data labelled as \textit{Ann. Phys. 2018} is the data published in \cite{Dudal:2018cli} and is, in all cases, plotted as a function of the improved momentum $\hat{p}$. 
The data referred as \textit{New Calc.}
is represented as a function of the improved momentum $\hat{p}$ (top-left), as a function of the lattice momentum $p$ (top-right) 
and as a function of the lattice momentum after a linear extrapolation in $p^{[4]}$ (bottom).
The \textit{New Calc.} data includes momenta of type $(n, \, 0, \, 0, \, 0)$,  $(n, \, n, \, 0, \, 0)$,  $ (n, \, n, \, n, \, 0)$,  $(n, \, n, \, n, \, n)$
and an average over the momenta with the same set of lattice invariants $p^2$, $p^{[4]}$, $p^{[6]}$ and $p^{[8]}$ 
was performed before doing the ensemble average. 
The data \textit{New Calc. + H4 Ext.} is the same as  \textit{New Calc.} after a linear extrapolation in $p^{[4]}$.
Recall  that for the class of  momenta considered, that involves a single momentum scale,
the Landau gauge condition $p \cdot A(p) \approx 0$ is verified independently of the type of momentum considered, i.e. independently of 
using $p$ or $\hat{p}$. Indeed, for the class of momentum configurations considered, that have a unique scale, the gauge condition translates
into a condition on the sum of the gauge field components.

The plots in Fig. \ref{fig:LatContBas80} report on the new calculation of the gluon propagator, exploring the H(4) invariants.
The top left plot shows that the data named \textit{New Calc.}  agrees well with the reference data but has smaller statistical errors as expected. Therefore, it is worthwhile to take averages over the lattice data with the same set 
of scalar invariants. 

However, there is a significant difference between $D(\hat{p}^2)$,
compared to $D(p^2)$; see top-right plot. The difference is relevant  for $p \gtrsim 1.5$ GeV and is clearly visible for $p \gtrsim 2$ GeV. 
The linear extrapolation in $p^{[4]}$ of the \textit{New Calc.} data returns a propagator that essentially reproduces the $D(\hat{p}^2)$ 
named \textit{Ann. Phys. 2018}. Note however that there are small deviations between the two sets of results.
The good agreement of these two estimates for $D(p^2)$ gives us confidence that the procedure devised in \cite{Leinweber:1998uu} and
used in many calculations for the gluon propagator, provides a proper estimation of $D(p^2)$.  

The data in  Fig. \ref{fig:LatContBas80} also shows that the use of the $H(4)$ extrapolation has advantage over the momentum cuts, providing
a larger number of data points over a large range of momenta. Recall that
the extrapolation does not work for low and large momenta, where the number of data points with the same $p^2$ but different
$p^{[4]}$ is not enough to allow for a reliable estimation of the form factor at $p^{[4]} = 0$. For momenta above $\sim 5$ GeV the linearly
extrapolated data and our reference data are no longer compatible within one standard deviation.

\subsection{On the completeness of the  tensor bases \label{Sec:R}}

\begin{figure}[t] 
   \centering
   \includegraphics[width=3.2in]{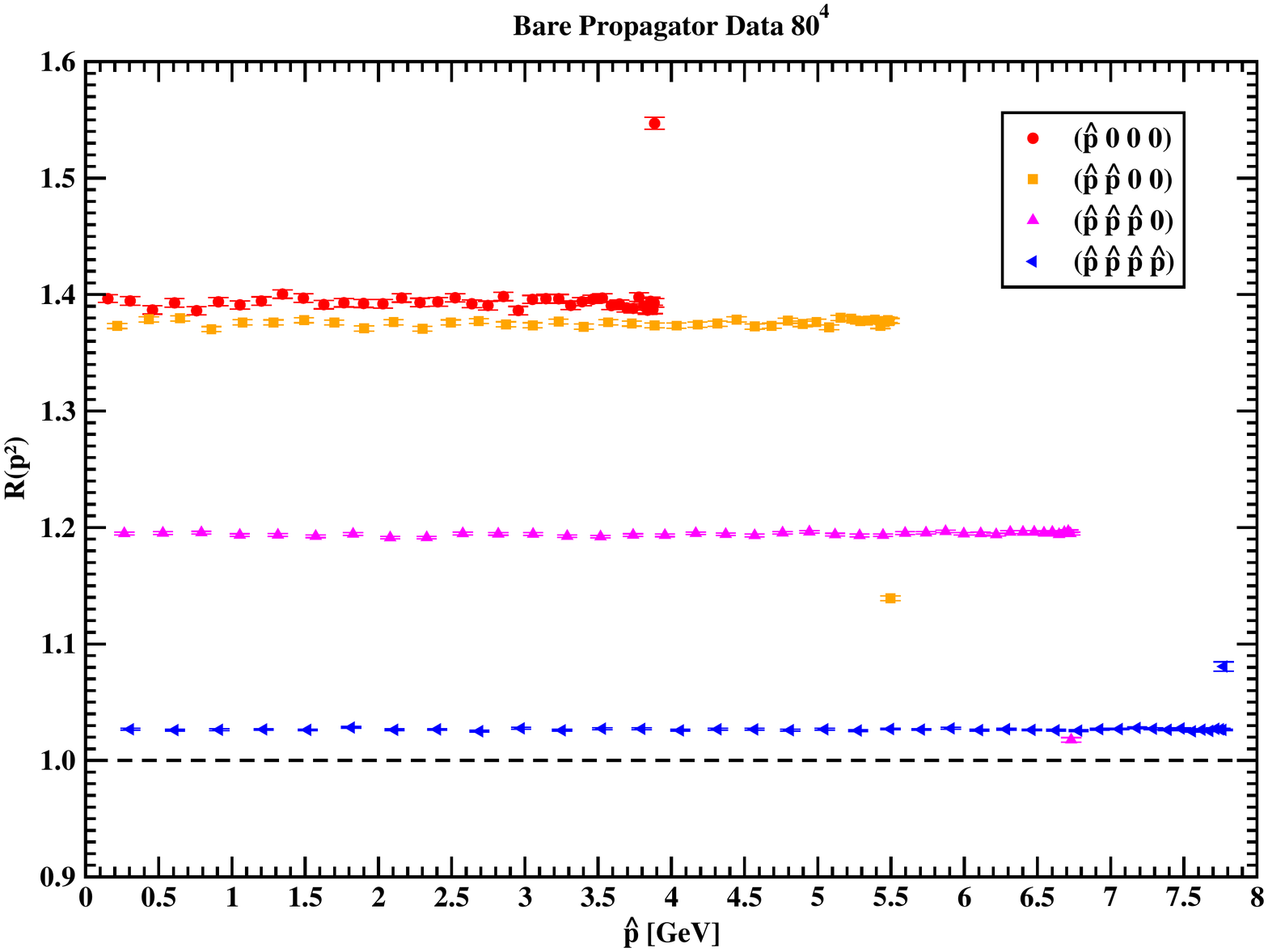}  ~
   \includegraphics[width=3.2in]{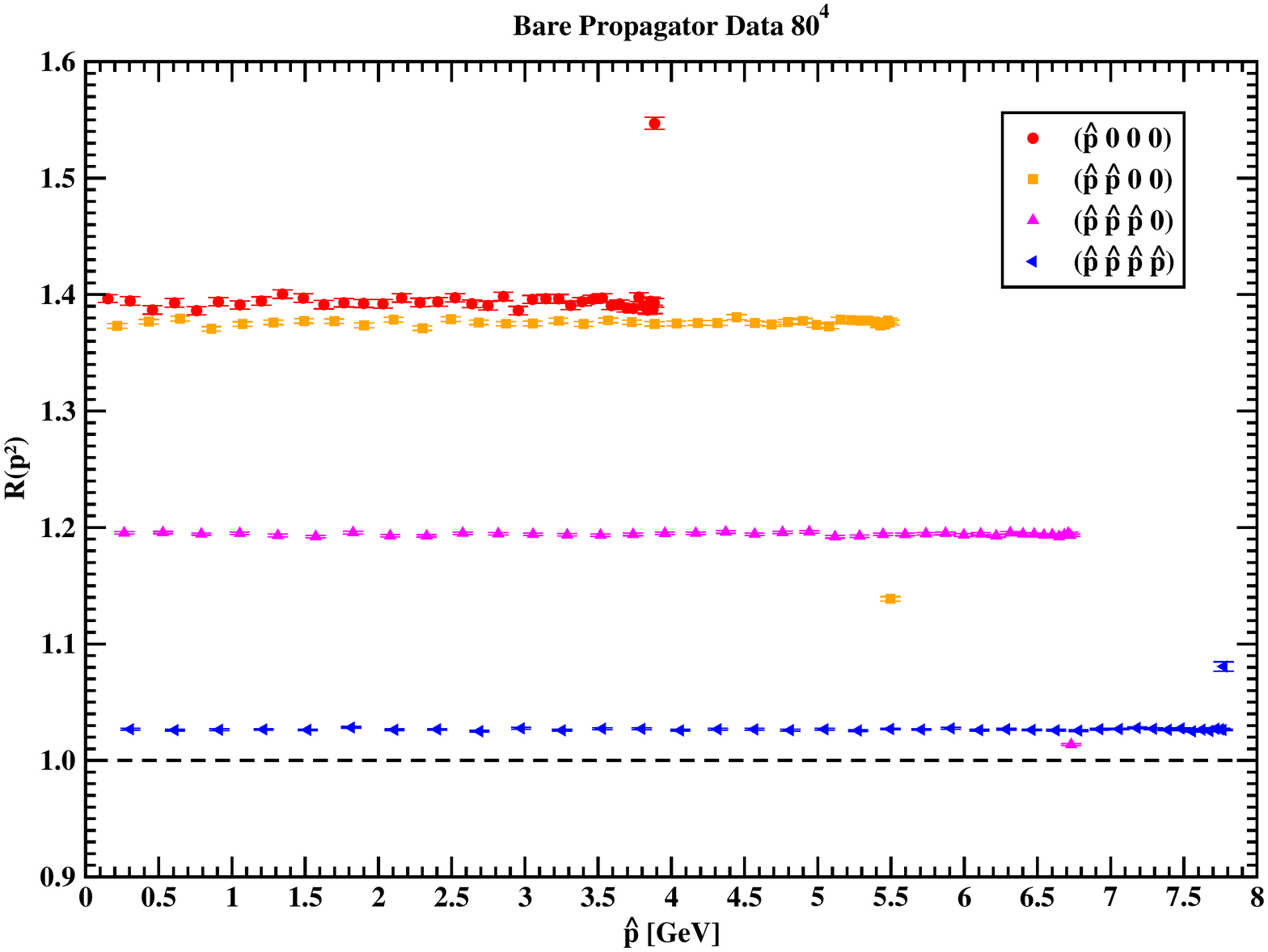} 
   \caption{$R$ for different classes of momenta that are defined by a single momentum scale $p$ as a function of the improved momenta when using the
                 continuum tensor basis (left) and for the extended tensor basis defined in Eq. (\ref{Eq:LandauLattProp}) (right). As discussed in the main text,
                 for diagonal momenta and for the extended tensor basis, it is not always possible to measure the full set of form factors.}
   \label{fig:R_LatContBas80}
\end{figure}

Let us now investigate how faithful is the description of the lattice Landau gauge gluon propagator when 
the continuum tensor basis of Eq. (\ref{Eq:ContProp}) is used. Due to the breaking of the rotational  symmetry one expects
deviations relative to the continuum tensor basis. The deviations can be tested through the ratio
\begin{equation}
   R = \frac{\sum_{\mu\nu} \left| \mbox{Tr} \, D^{Lat}_{\mu\nu} (p) \right| }{\sum_{\mu\nu} \left| \mbox{Tr} \, D^{rec}_{\mu\nu} (p) \right| } 
   \label{Eq:Def_R_Latt}    \  ,  
\end{equation}
where $D^{Lat}_{\mu\nu} (p)$ is the lattice propagator as given by the simulation, i.e. as measured from Eq. (\ref{Eq:DefProp}),
and $D^{rec}_{\mu\nu} (p)$ is the reconstructed propagator using the form factor $D(p^2)$ and assuming the tensor structure as in Eq. (\ref{Eq:ContProp}). 
A complete tensor basis should describe the lattice propagator with great accuracy and 
the corresponding $R$ should be one.

In Fig. \ref{fig:R_LatContBas80}  the ratio $R$ is reported for the class of momenta which have a single momentum scale $p$ and for the continuum tensor basis. 
Note that in the plots the data for $R$ is reported as a function of the improved momentum $\hat{p}$. 
There is no significative difference on $R$ if one uses either the improved momenta $\hat{p}$ or the lattice momenta $p$. 
As seen, $R$ deviates significantly from unity and can reach values just below 
$\sim 1.4$ for on-axis and $(p, \, p, \, 0, \, 0)$ type of momenta.
For the $(p, \, p, \, p, \, 0)$ class of momenta $R$ drops to $\sim 1.2$ and for diagonal momenta it is close to the ideal value.

\begin{figure*}[t] 
   \centering
   \includegraphics[width=3.3in]{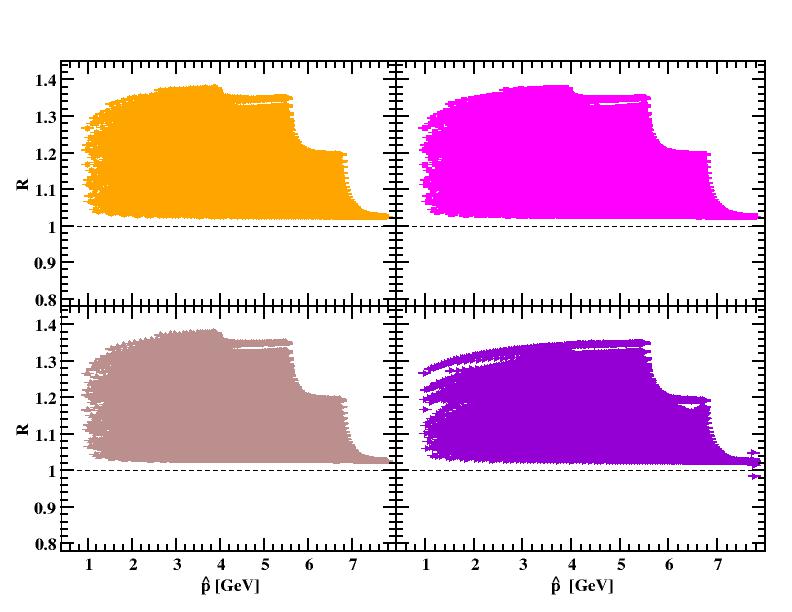}  
   \includegraphics[width=3.3in]{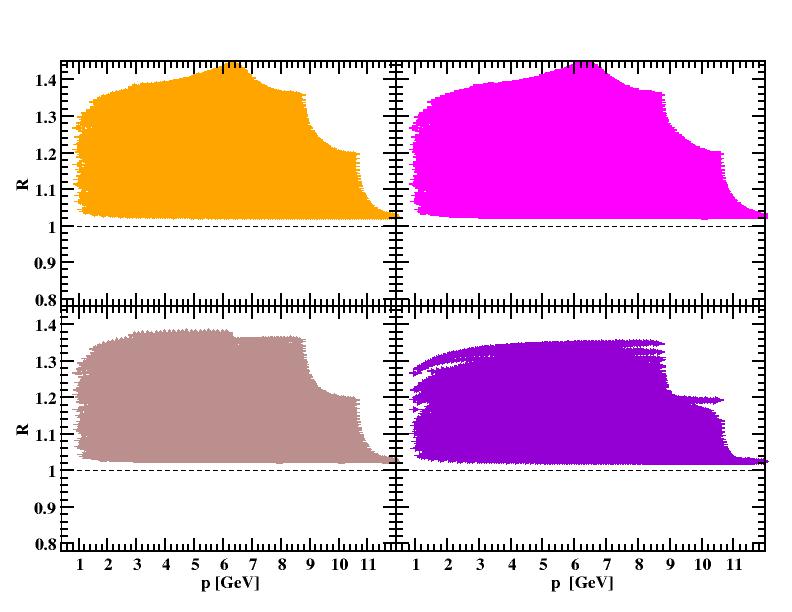}  \\
   \includegraphics[width=3.3in]{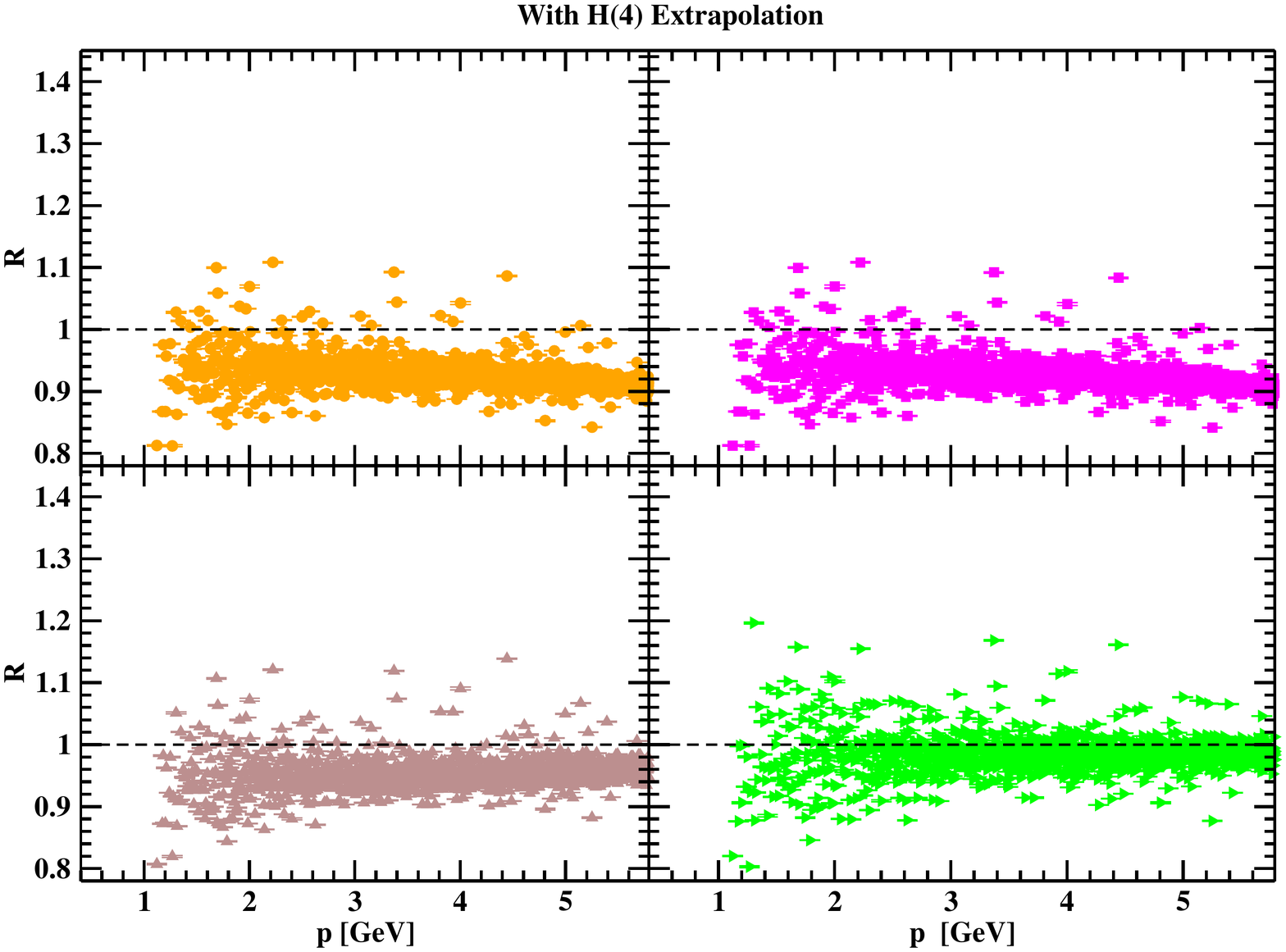}
   \caption{$R$ for the continuum basis (\ref{Eq:ContProp}) (top left), for the basis with two form factors (\ref{eq:lattice_continuum_basis}) (top right), 
                 for the extended basis (\ref{eq:partial_lattice_basis}) that requires three form factors (bottom left) and for the enlarged basis  
                 (\ref{Eq:LandauLattProp0}) that calls for five form factors (bottom right).
                 The three sets of plots use the improved momentum $\hat{p}$ (top left), the lattice momentum $p$ (top right) 
                 and the lattice momentum after $H(4)$ extrapolation.}
                 \label{fig:LatRecAll}
\end{figure*}

Let us now see if the use of larger tensor bases can help improving $R$ and provide a more faithful description of the lattice data. In Fig. \ref{fig:LatRecAll}
we report on $R$, for all available momenta where one can measure the corresponding basis form factors, 
computed using the continuum basis (\ref{Eq:ContProp}) (top left), the basis with two form factors (\ref{eq:lattice_continuum_basis}) (top right), 
the extended basis (\ref{eq:partial_lattice_basis}) that requires three form factors (bottom left) and for the enlarged basis
(\ref{Eq:LandauLattProp0}) that calls for five form factors (bottom right), for each of the set of plots. 
The three sets of plots the data show $R$ as a function of the improved momentum $\hat{p}$ (top left), 
the lattice momentum $p$ (top right) and the lattice momentum after $H(4)$ extrapolation\footnote{We postpone the computation of the form factors for the
extended tensor basis to Sec. \ref{Sec:GProp_Ext}.}. 
Recall that the $H(4)$ extrapolation is not possible at  small and high momenta. 
This is the rationale why the extrapolated data in Fig. \ref{fig:LatRecAll} includes only momenta with $p \lesssim 6$ GeV.
As seen, the results for $R$ are slightly improved, more in the reduction of the interval of values taken than
by bringing $R$ towards its optimal value, as one consider larger tensor bases. The worst scenario occurs when the lattice data is described in terms of the naive
lattice momentum $p$. The use of the improved lattice momentum $\hat{p}$ reduces the spread of $R$ but it does not bring it closer to one. On the other
hand, the use of the naive lattice momentum combined with the $H(4)$ extrapolation has a large impact on $R$, resulting in $R$ values that are clearly 
close to its optimal value. Recall that the errors associated with the extrapolation do not rely on the bootstrap method and are underestimated. 
Moreover, looking at the extrapolation data for $R$, it is for the enlarged basis (\ref{Eq:LandauLattProp}) that the extrapolation returns $R$ values closer
to the unity. 

In what concerns the faithfulness of the description of the Landau gauge gluon propagator, the results of Fig. \ref{fig:LatRecAll} suggest that one should use large
tensor bases combined with $H(4)$ extrapolations. Eventually, at least in the infinite limit statistics, a sufficiently large tensor basis will exempt the extrapolation.
Notice, however, that the results 
summarised in Figs. \ref{fig:R_LatContBas80} and \ref{fig:LatRecAll} have little to do with the quality of the measured form factor $D(p^2)$. 
They are essentially a test of the (un)completeness of the lattice tensor basis. 

The results summarised  in Fig. \ref{fig:R_LatContBas80} show that the continuum tensor structure  is not suitable to describe, with precision, the lattice propagator. 
The relative lack of precision on the reconstruction of the lattice propagator can be due to large corrections to the form factor $D(p^2)$ (previous studies suggest 
that this does not occur), that large lattice corrections should occur mainly in the non-diagonal components or that the definition used for the gluon field needs 
to be changed. We have tested the later case by
improving the orthogonality of the gluon field, i.e. replacing 
$A_\mu (p)$ by
\begin{equation}
   A^{(ort)}_\mu (p) = \left( \delta_{\mu\nu} - \frac{p_\mu p_\nu}{p^2} \right) A_\mu (p) \ .
   \label{Eq:Aorto}
\end{equation}
The corresponding analysis of the data with this new definition for the gluon field does not change neither the propagator nor $R$.
This  procedure tests the orthogonality condition on the lattice and suggests that orthogonality is well satisfied by the conventional 
definition of the gluon field.

Our analysis of $R$
shows that, for the class of momenta considered, the extended tensor is not yet a complete basis in the sense that it is not able to provide a 
faithful reconstruction of the lattice propagator. 
Probably this can be achieved by considering larger tensor  bases. However, see the discussion below,   the statistical errors
for the form factors $G$ and $I$ are large and oftentimes these form factors are compatible, within one standard deviation, with zero. 
This prevent us from considering other larger tensor bases with larger sets of form factors.

\subsection{Results for the largest tensor basis \label{Sec:GProp_Ext}}

Let us now look at the form factors $E(p^2)$, $F(p^2)$, $G(p^2)$, $H(p^2)$ and $I(p^2)$ that appear in the extended tensor basis defined in
Eq. (\ref{Eq:LandauLattProp}).
Only those kinematical configurations where the full set of form factors, as given by Eqs. (\ref{Eq:Proj_E}) - (\ref{Eq:Proj_I}), can be
accessed will be considered.

The extended basis given in Eq. (\ref{Eq:LandauLattProp}) reduces to the continuum tensor basis (\ref{Eq:ContProp}) if the relations
\begin{equation}
D(p^2) = E(p^2) = - p^2 \, F(p^2) = - p^2 \, H(p^2) \qquad\mbox{ and }\qquad G(p^2) =  I(p^2) = 0 \label{Eq:Ext2ContBase}
\end{equation}
are verified. These conditions, together with the comparison with the standard computation of $D(p^2)$, will be used to benchmark the measurement of
the form factors. Similar relations can be defined to analyse the other tensor bases mentioned previously.
  
\begin{figure}[t] 
   \centering
   \includegraphics[width=3.45in]{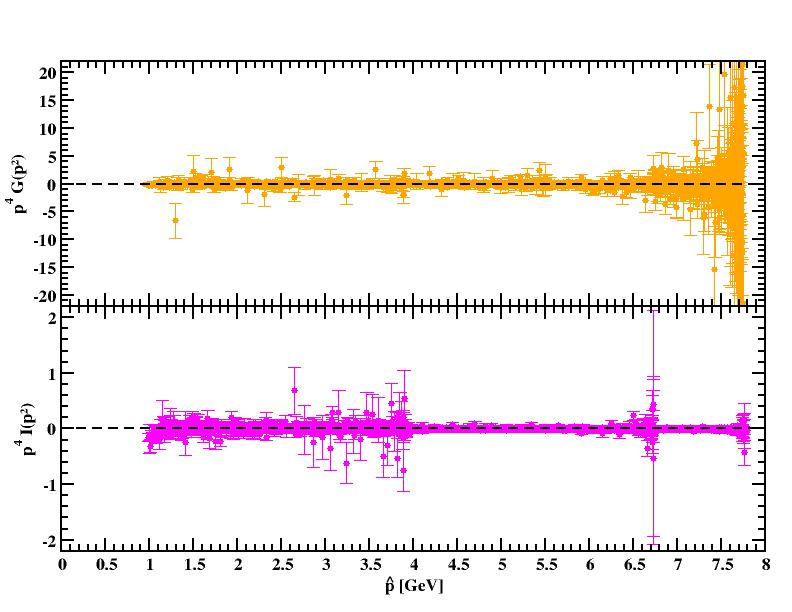}  ~
   \includegraphics[width=3.45in]{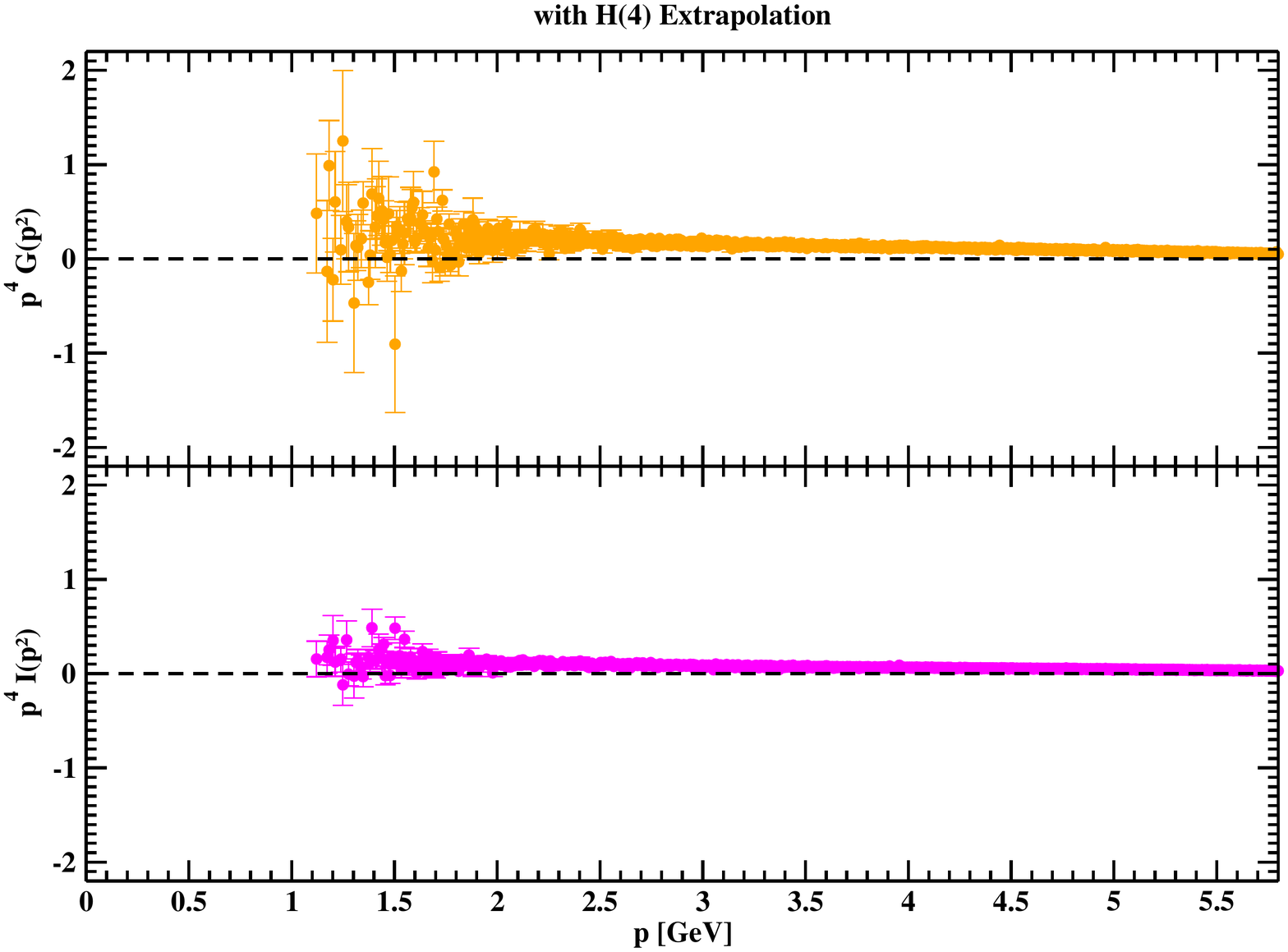}  
   \caption{$p^4 \, G(p^2)$ and $p^4 \, I(p^2)$ (dimensionless 
      quantities) as function of the improved momentum (left)
        and as function of the lattice momentum after $H(4)$ extrapolation (right).
     The large fluctuations observed at high $\hat{p}$ for $G(p^2)$ are associated with small values of $\Delta_1$.}
   \label{fig:G_and_I_Pimp80}
\end{figure}

In Fig. \ref{fig:G_and_I_Pimp80} the form factors $p^4 \, G(p^2)$ and $p^4 \, I(p^2)$ are reported  as a function of the improved momentum (left plot) for all the kinematical configurations, and of the lattice momentum $p$, after $H(4)$ extrapolation (right plot).
$p^4 I(p^2)$ is compatible with zero for the two cases considered while $p^4 G(p^2)$ is compatible with zero, in the full range of momenta,
only when the computation uses the improved momentum. Indeed, $p^4 G(p^2)$ deviates slightly from zero when the data is represented in terms 
of the lattice momentum, after the linear extrapolation in $p^{[4]}$ at the smallest momenta. 
Once more, we recall the reader that the evaluation of the statistical errors with the extrapolation do not use the  bootstrap method and rely only
on the error evaluation of the linear regression which are underestimated.

The fluctuations of the data are smaller for $p^4 I(p^2)$ when compared to the fluctuations of $p^4 G(p^2)$. The pattern of the fluctuations is as expected 
given that $G(p^2)$ mixes with $E(p^2)$ and $F(p^2)$ that, in principle, are the most relevant form factors for the Landau gauge gluon propagator. 
As seen below, the data confirms the relative importance of the form factors.
Moreover, Fig. \ref{fig:G_and_I_Pimp80} suggests also that, within the statistical precision of the current simulation, it is difficult to go beyond the extended 
tensor basis Eq. (\ref{eq:partial_lattice_basis}) to describe the lattice gluon propagator in the Landau gauge.

The small observed values for the form factors represented on Fig. \ref{fig:G_and_I_Pimp80} suggest that $p^4 I(p^2) \approx p^4 G(p^2) \approx 0$ for
all  momenta. In this case, 
it follows from the discussion in Sec. \ref{Sec:orto_gen}, see Eq. (\ref{Eq:TransvLatt2}), that
the orthogonality condition for a general kinematical configuration implies for the lattice data
\begin{equation}
    E(p^2)   +  F(p^2) \, p^2  = 0  \ ,
\end{equation}
and, from Eq. (\ref{Eq:Cont_D_ExtBas_Real}), that the continuum form factor $D(p^2) = E(p^2)$. Similar relations also hold for on-axis momenta, see Eqs.
(\ref{Eq:Mais_uma_OnAxis}) and (\ref{Eq:RefOnAxis}). For diagonal momenta, if the relation $H(p^2) = F(p^2)$ applies, then once more
$D(p^2) = E(p^2)$, see Eqs. (\ref{Eq:RefDiagonalDiagonal}) and (\ref{Eq:OrtoLattDiag}).

\begin{figure}[t] 
   \centering
   \includegraphics[width=3.45in]{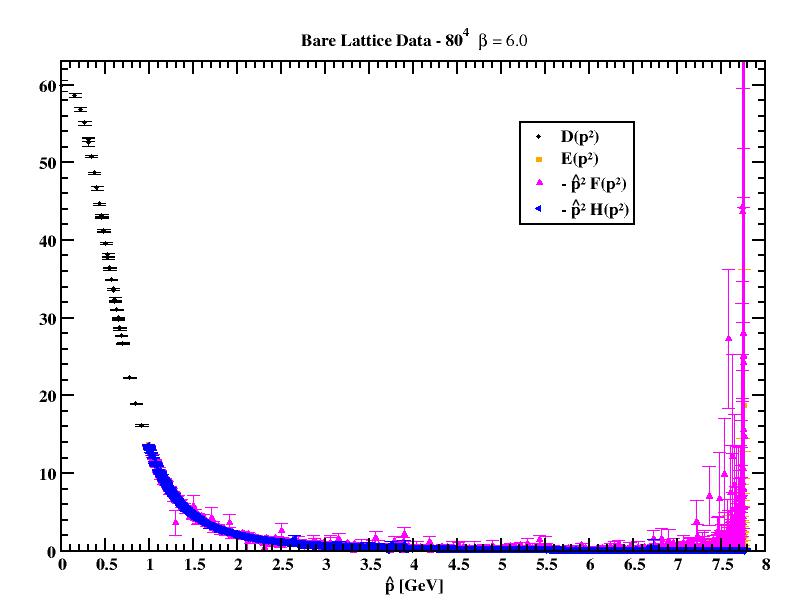}  ~   
   \includegraphics[width=3.45in]{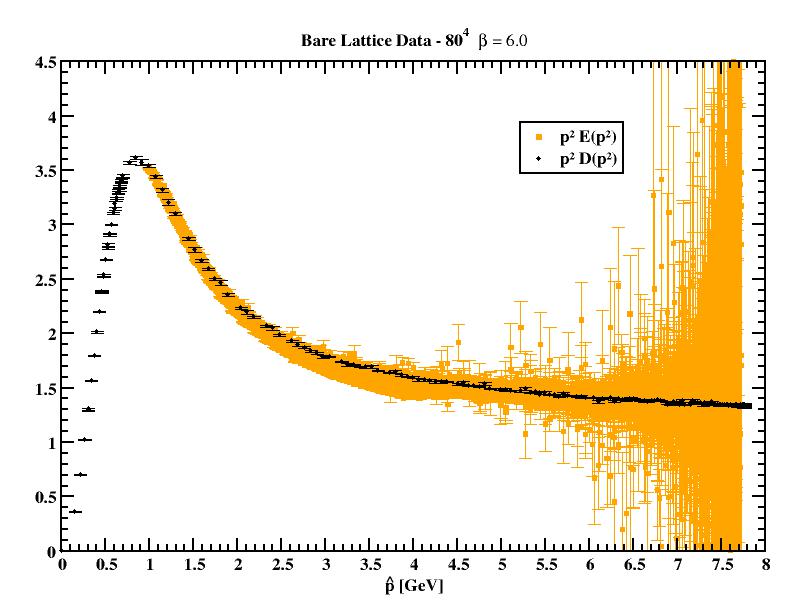}  \\
   \includegraphics[width=3.45in]{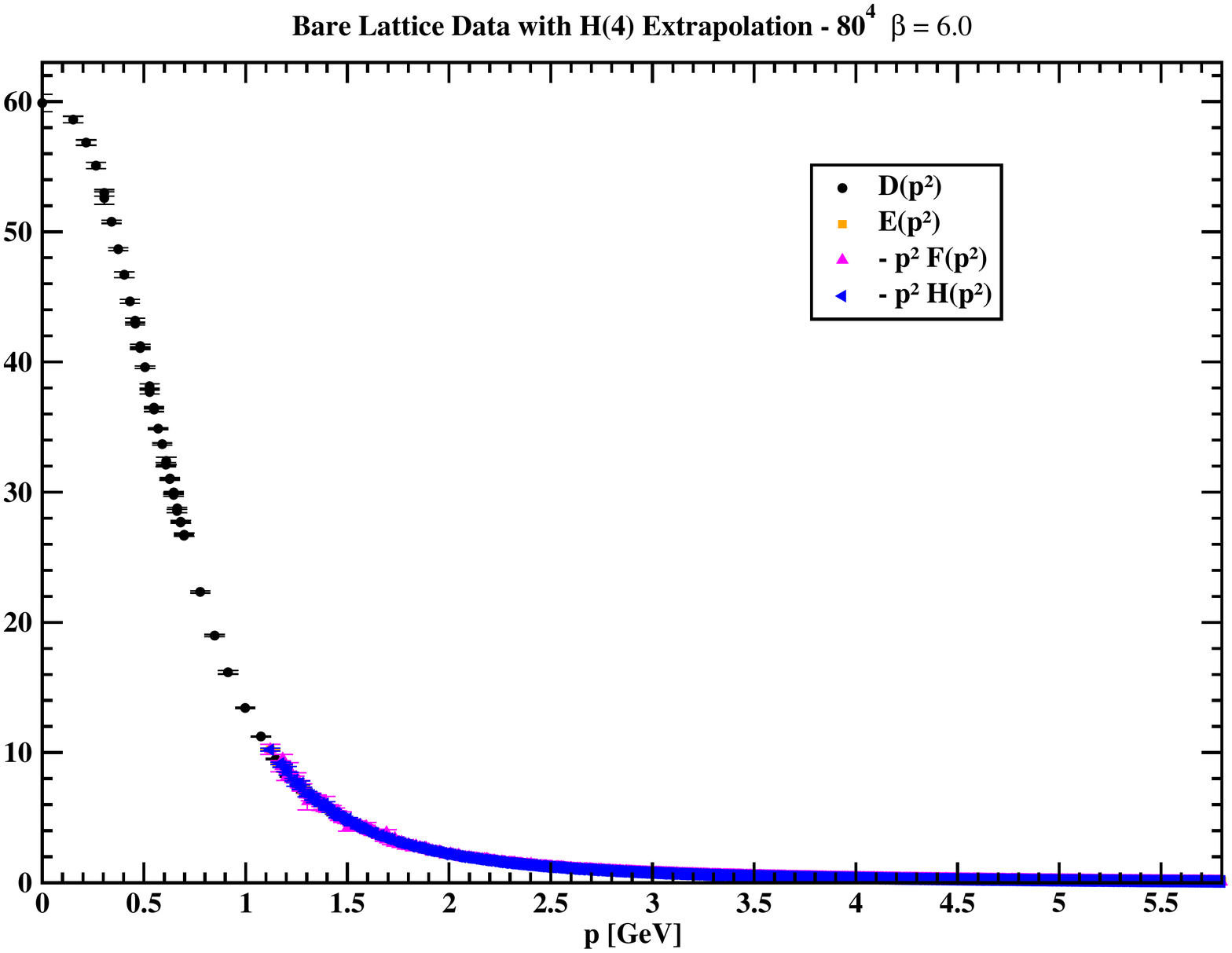}  ~
   \includegraphics[width=3.45in]{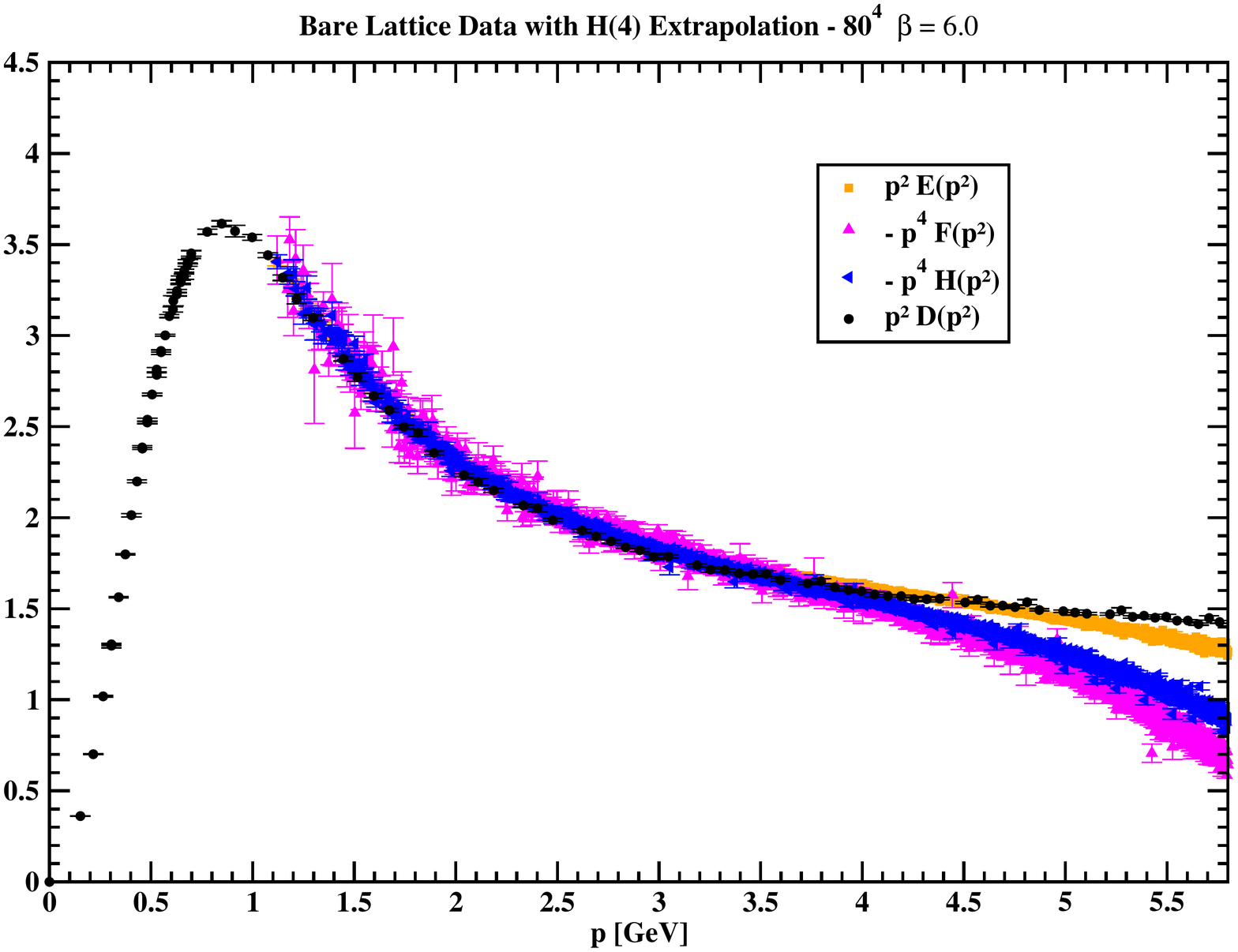} 
   \caption{Form factors (dimensionless units) as a function of the improved momentum (upper) and as a function of the lattice momentum after $H(4)$ 
                 extrapolation (bottom). The function $D(p^2)$ is plotted always as a function of the improved momentum $\hat{p}$.
                 The left plots refers to the form factors that, in the continuum, should be equal, while the right plots refers to the same form factors after
                 multiplication by momentum squared. 
                 The right plots do not include all the form factors as the lattice data associated with those not represented on the right plots have large statistical errors.
                 The inclusion of this data does not add any useful information and it overshadows the plots reading.}
   \label{fig:D_E_F_H_L80}
\end{figure}

In Fig. \ref{fig:D_E_F_H_L80} we show the lattice data for all form factors and test the relations (\ref{Eq:Ext2ContBase}) that are required to reproduce the 
continuum tensor basis structure. In all cases, the quantities represented are  dimensionless. 
The data is given in terms of the improved momentum $\hat{p}$ in the upper plots and, in the bottom plots, as a function of the lattice momentum $p$ after
performing the linear extrapolation in $p^{[4]}$. 
Recall that the form factor $D(p^2)$ obtained using the continuum tensor basis and momentum cuts, is always given as a function of the improved momentum. 
The combinations reported in Fig. \ref{fig:D_E_F_H_L80}  are such that in the continuum limit they  should all become equal. Any deviation between 
the form factors is a manifestation of finite volume and/or finite spacing effects. For the gluon dressing function seen as a function of
the improved momentum $\hat{p}$, $p^2 E(p^2)$ follows the behaviour of the full data set observed in Fig. \ref{fig:FF_Cont_Basis_NoBin} and, therefore,
it seems that the use of extended tensor basis does not disentangle the lattice artefacts. These are pruned when we consider the lattice data as a function of the
naive momentum after performing the $H(4)$ extrapolation.
In general, the form factors are compatible with each other, within one standard deviation. Together with the observations that $G(p^2)$ and $I(p^2)$ are essentially 
vanishing or quite small functions of the momenta,  these results imply that  the simulation reproduces essentially the continuum relations between the form factors. 

This can also be viewed as an indication that the tensor structure of the continuum propagator, or the Slavnov-Taylor identity for the gluon, 
holds for the lattice data. This result confirms the conclusion reached in \cite{Oliveira:2012eh} that for the simulation under discussion, 
the lattice results reproduce continuum physics.

The data in Fig.  \ref{fig:D_E_F_H_L80} also shows an hierarchy on the quality of the lattice data associated with each of the form factors 
for the extended tensor basis (\ref{Eq:LandauLattProp}). $E(p^2)$ has the smallest statistical errors, while $- \, p^2 F(p^2)$ 
and $- \, p^2 H(p^2)$ have the largest statistical errors. From the point of view of accessing useful information on the gluon propagator, it is $E(p^2)$ 
that delivers the best Monte Carlo signal.

\subsection{Comparing The Tensor Bases}

\begin{figure*} 
   \centering
   \includegraphics[width=6.1in]{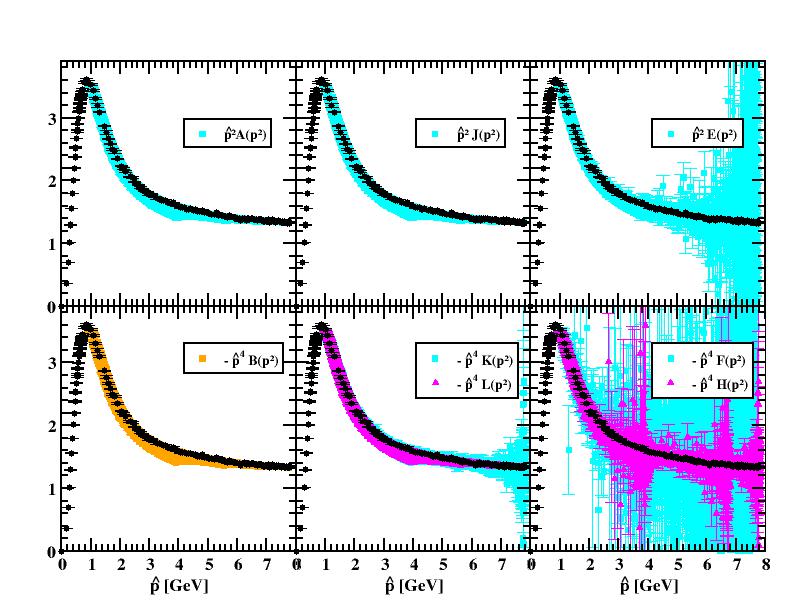}  \\
   \vspace{-0.3cm}
   \includegraphics[width=6.1in]{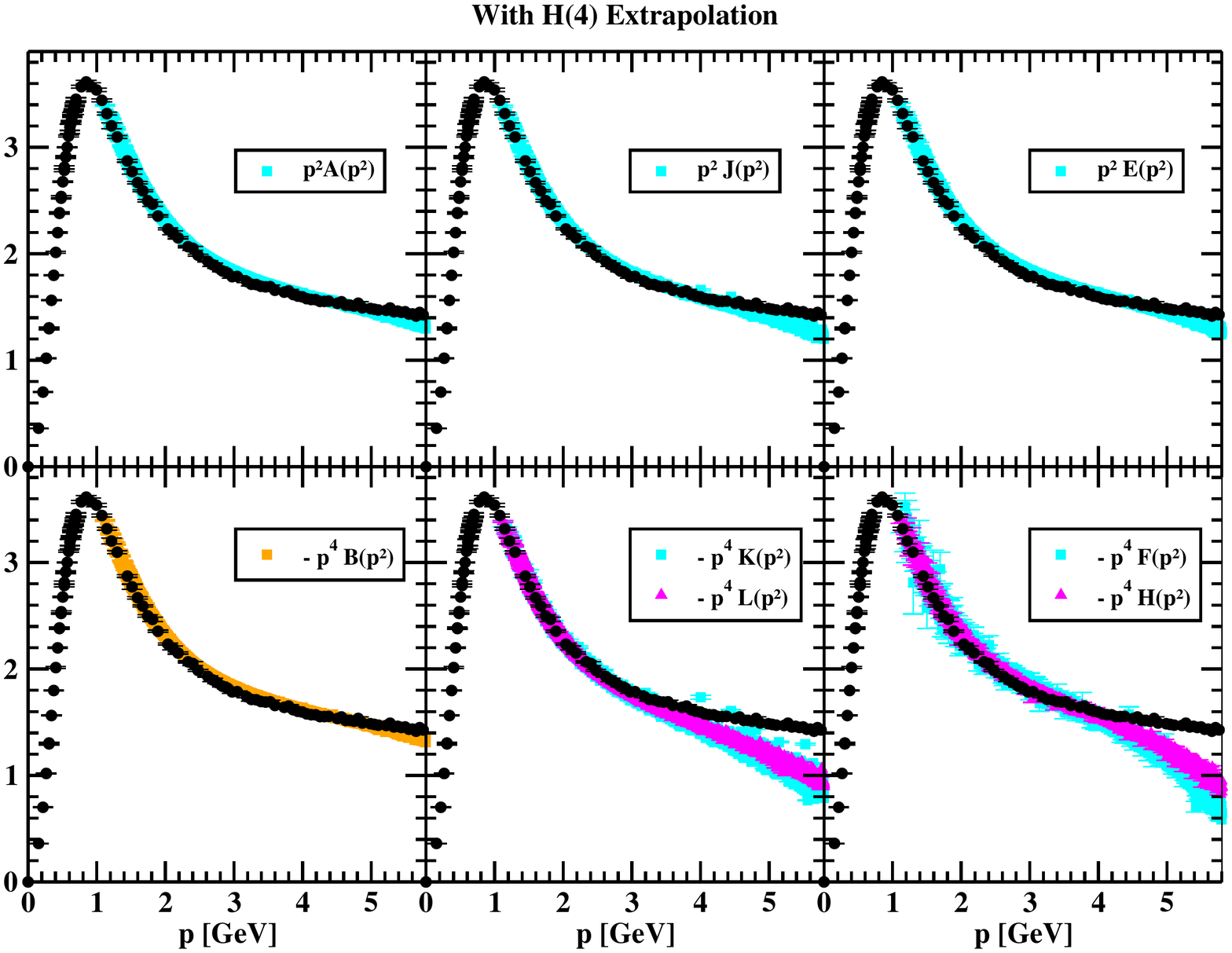} 
   \vspace{-0.5cm}
   \caption{Dimensionless form factors for all the different tensor basis, compared to $D(p^2)$ (black circles in all the figures). 
                The top line shows the form factors as a function of the lattice improved momentum. The bottom line are the same 
                results but using the lattice momentum followed by a linear $p^{[4]}$ extrapolation. The
                $D(p^2)$ data is given as a function of the improved momentum.}
   \label{fig:FF_All_Basis}
\end{figure*}

So far, we have observed that by enlarging the tensor basis the accuracy of the description of the Landau gauge lattice gluon propagator increases slightly
and that the linear $H(4)$  extrapolation has a major impact on the faithfulness and on the handling of the lattice artefacts.
In this section, we look at the results obtained with the (\ref{eq:lattice_continuum_basis}) and 
the (\ref{eq:partial_lattice_basis}) bases, and compare them to the standard approach to the gluon propagator and with the results of 
the extended tensor basis (\ref{Eq:LandauLattProp}). 
The analysis of the various bases requires the computation of the projection operators that are similar to those already quoted in the main text.

In Fig. \ref{fig:FF_All_Basis} we show the form factors associated with the various tensor bases.
The form factors are compared to $D(p^2)$ computed in \cite{Dudal:2018cli} given as a function of the improved lattice momentum. 
The results reported in Fig. \ref{fig:FF_All_Basis} use either the lattice improved momentum (top plots) or the lattice momentum after a linear extrapolation
in $p^{[4]}$ of the form factors (bottom plots). For each set of plots, the form factors associated to the metric tensor are reported in the top line, 
while the remaining ones appear in the bottom line. For comparison, we always show $D(p^2)$, as a function of the improved
momentum, for the momentum configurations that verify the cuts mentioned previously.

If one describes the form factors in terms of the improved momentum $\hat{p}$, 
the data for all the form factors seems to follow the functional dependence observed in $D(\hat{p}^2)$ and, in this sense,
one can claim that they reproduce $D(p^2)$. In general, $\hat{p}^2 A(\hat{p}^2)$, $\hat{p}^2 J(\hat{p}^2)$, $\hat{p}^4 B(\hat{p}^2)$, 
and $\hat{p}^4 L(\hat{p}^2)$ the structure observed in Fig.  \ref{fig:FF_Cont_Basis_NoBin} when considering the full
set of momenta is also observed in the top plots of Fig. \ref{fig:FF_All_Basis}. The analysis of the remaining
form factors, i.e.  $\hat{p}^2 E(\hat{p}^2)$, $\hat{p}^4 K(\hat{p}^2)$ , $\hat{p}^4 F(\hat{p}^2)$ and $\hat{p}^4 H(\hat{p}^2)$, is more difficult to disentangle as the statistical
errors are larger\footnote{Note that in Fig \ref{fig:FF_All_Basis} and in Fig. \ref{fig:D_E_F_H_L80} the range used in the 
vertical scale is not exactly the same.}. However, there is a difference on the data for 
$\hat{p}^2 A(\hat{p}^2)$, $\hat{p}^2 J(\hat{p}^2)$, $\hat{p}^4 B(\hat{p}^2)$, $\hat{p}^4 L(\hat{p}^2)$ form factors and 
$\hat{p}^4 K(\hat{p}^2)$, $\hat{p}^2 E(\hat{p}^2)$, $\hat{p}^4 F(\hat{p}^2)$ and $\hat{p}^4 H(\hat{p}^2)$. If the first set
is typically below the reference data, represented by $\hat{p}^2 D(\hat{p}^2)$, the later ones fluctuate around $\hat{p}^2 D(\hat{p}^2)$.

On the other hand, the description of the form factors using the naive lattice momentum combined with the $H(4)$ extrapolation
shows a much cleaner behaviour, with smaller statistical errors and with the various form factors reproducing the reference data up to
$\sim 4 $ GeV but not above it. Note also that the $H(4)$ extrapolated data does not produce a smooth curve but instead a
not so large band of values for each $p$. Once more, the extrapolation seems to be able to take into account and remove the lattice artefacts,
at least in a range of momenta.

\begin{figure*} 
   \centering
   \includegraphics[width=4in]{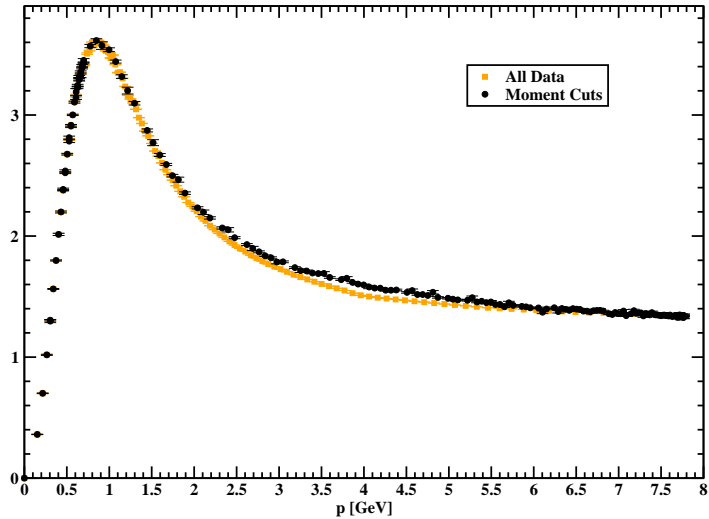} 
   \caption{The gluon dressing function $d(p^2) = p^2 D(p^2)$ as a function of the improved momentum for the continuum tensor basis for
                 the momentum configurations that verify the momentum cuts (back points) and all the kinematical configurations (orange points). 
                 The data for the full set is obtained by binning the lattice data as described in the main text.}
   \label{fig:FF_Cont_Basis_Bin2p5}
\end{figure*}

\begin{figure*} 
   \centering
   \includegraphics[width=6.1in]{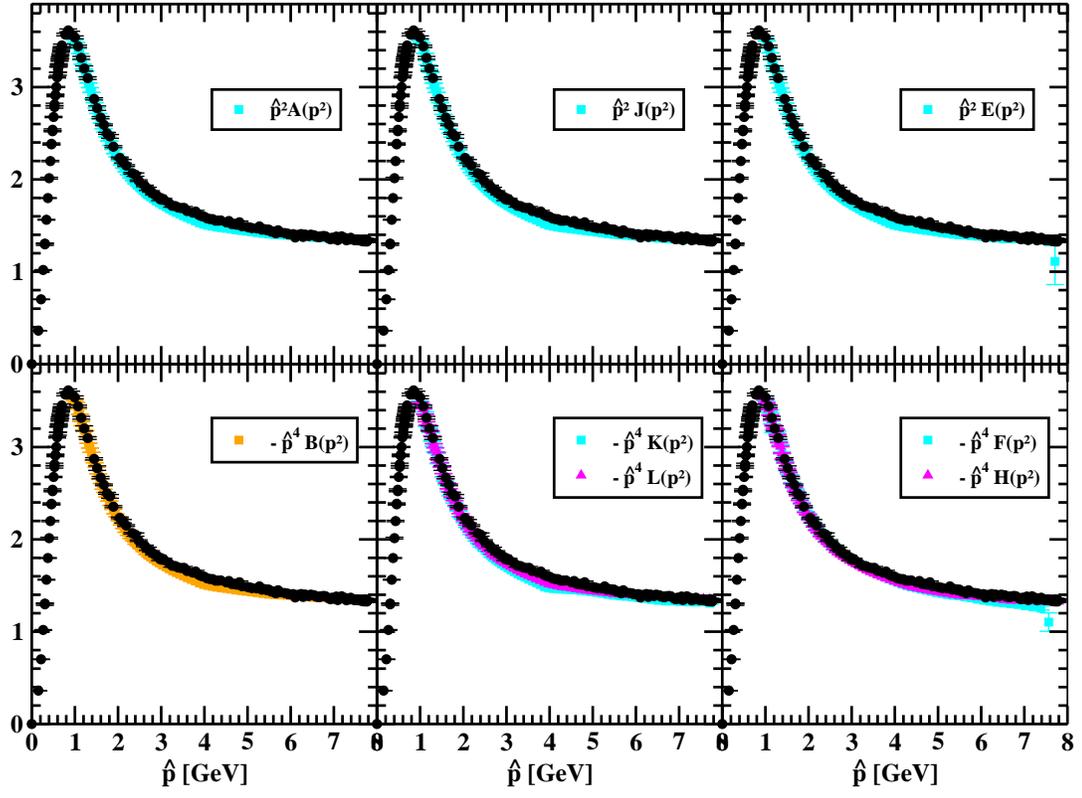}  \\
   \vspace{-0.3cm}
   \includegraphics[width=6.1in]{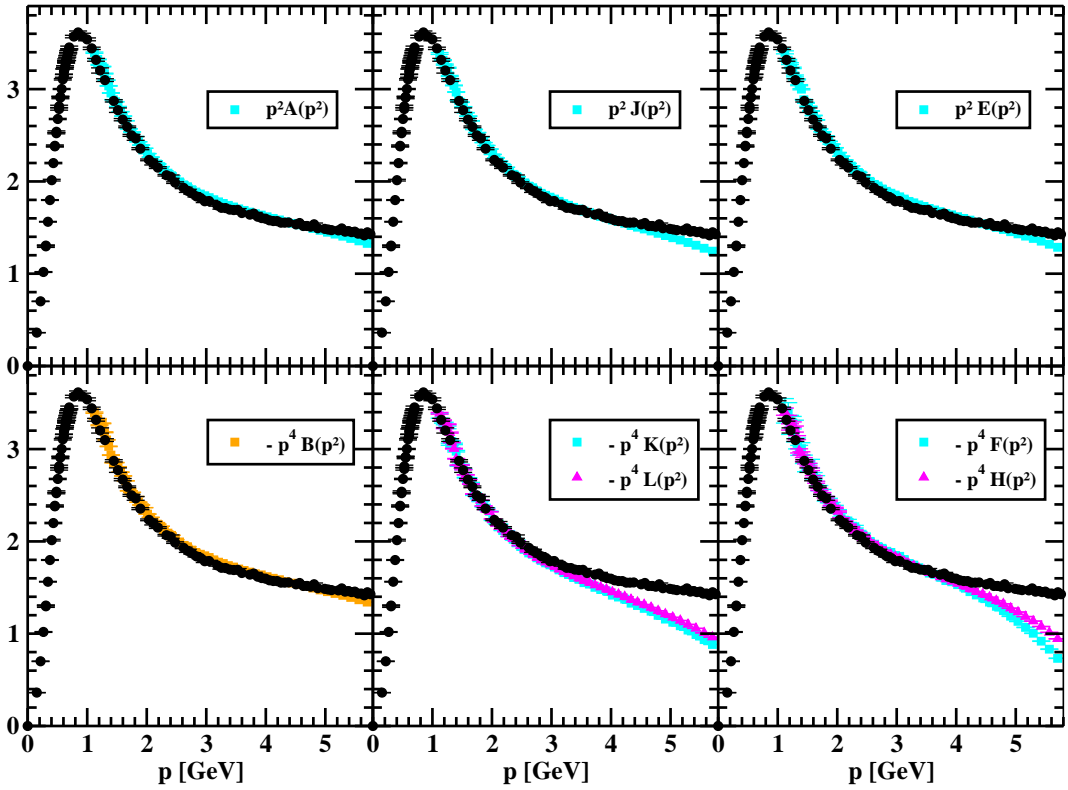} 
   \vspace{-0.5cm}
   \caption{The same as Fig. \ref{fig:FF_All_Basis} but after performing a binning on the horizontal axis using an interval of 2.5\%.}
   \label{fig:FF_All_Basis_Bin2p5}
\end{figure*}

At least partially, the problem of the noise level in the computation of the various form factors
can be overcome by recalling that there is an error on the scale setting of about 2.5\%. 
This 2.5\% ambiguity  can be used to define bins in the momentum and to replace all the data points in each bin 
by a weighted average of the data points. For each bin, we take the central value of the interval as the momentum.
The outcome of performing the binning can be seen in Figs. \ref{fig:FF_Cont_Basis_Bin2p5} and \ref{fig:FF_All_Basis_Bin2p5}.
In this way, one obtains smoother curves for all the form factors. In particular, in Fig. \ref{fig:FF_Cont_Basis_Bin2p5} that reports
the full data set for the continuum basis, the differences between considering the momentum cuts and the complete set of momentum
are clearly observed. These differences between the two data sets of momenta, i.e. the lattice artefacts, are clearly seen in the range
$\hat{p} \sim 2 - 5$ GeV.

\begin{figure*} 
   \centering
   \includegraphics[width=3.4in]{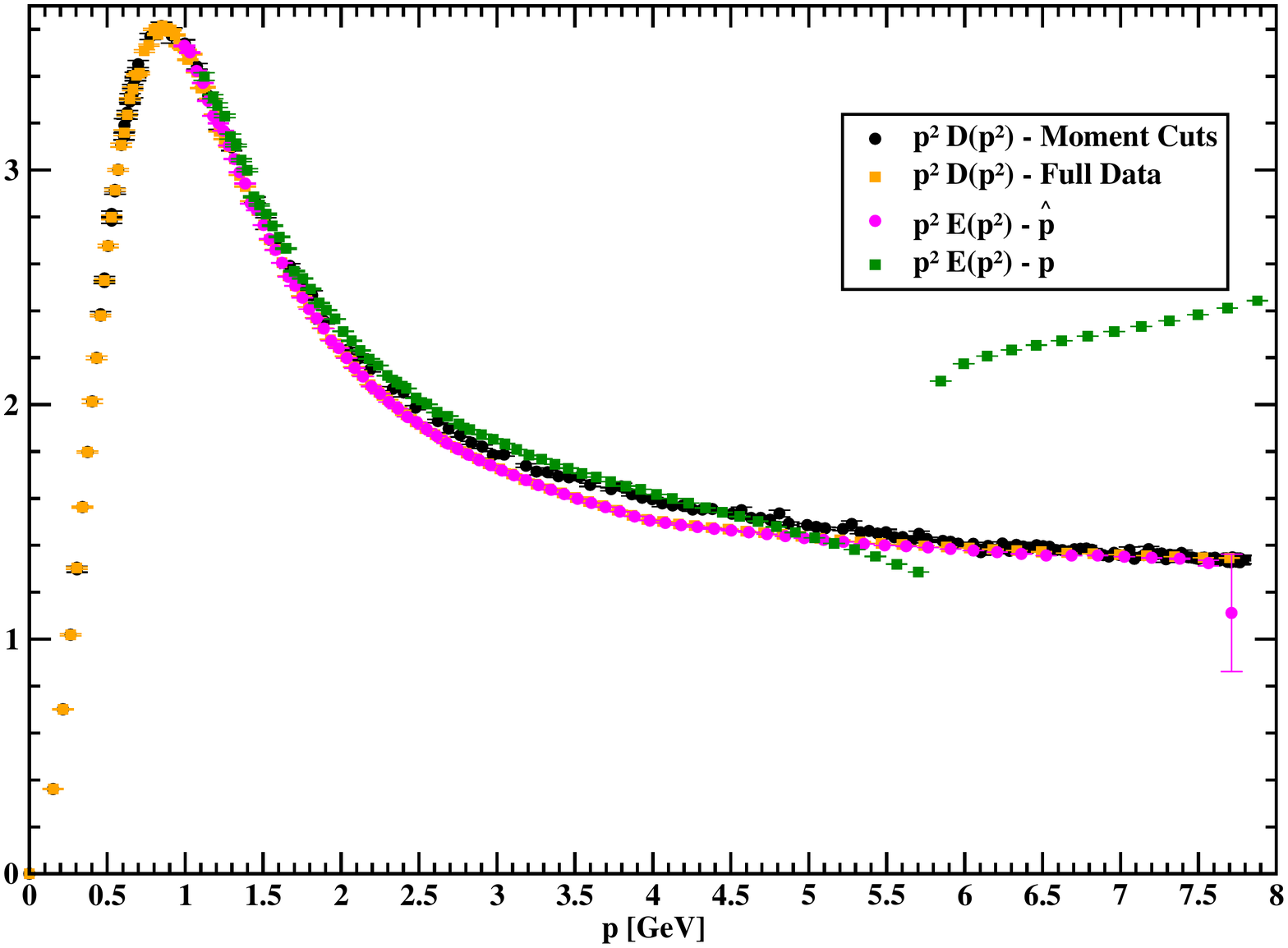} ~
   \includegraphics[width=3.4in]{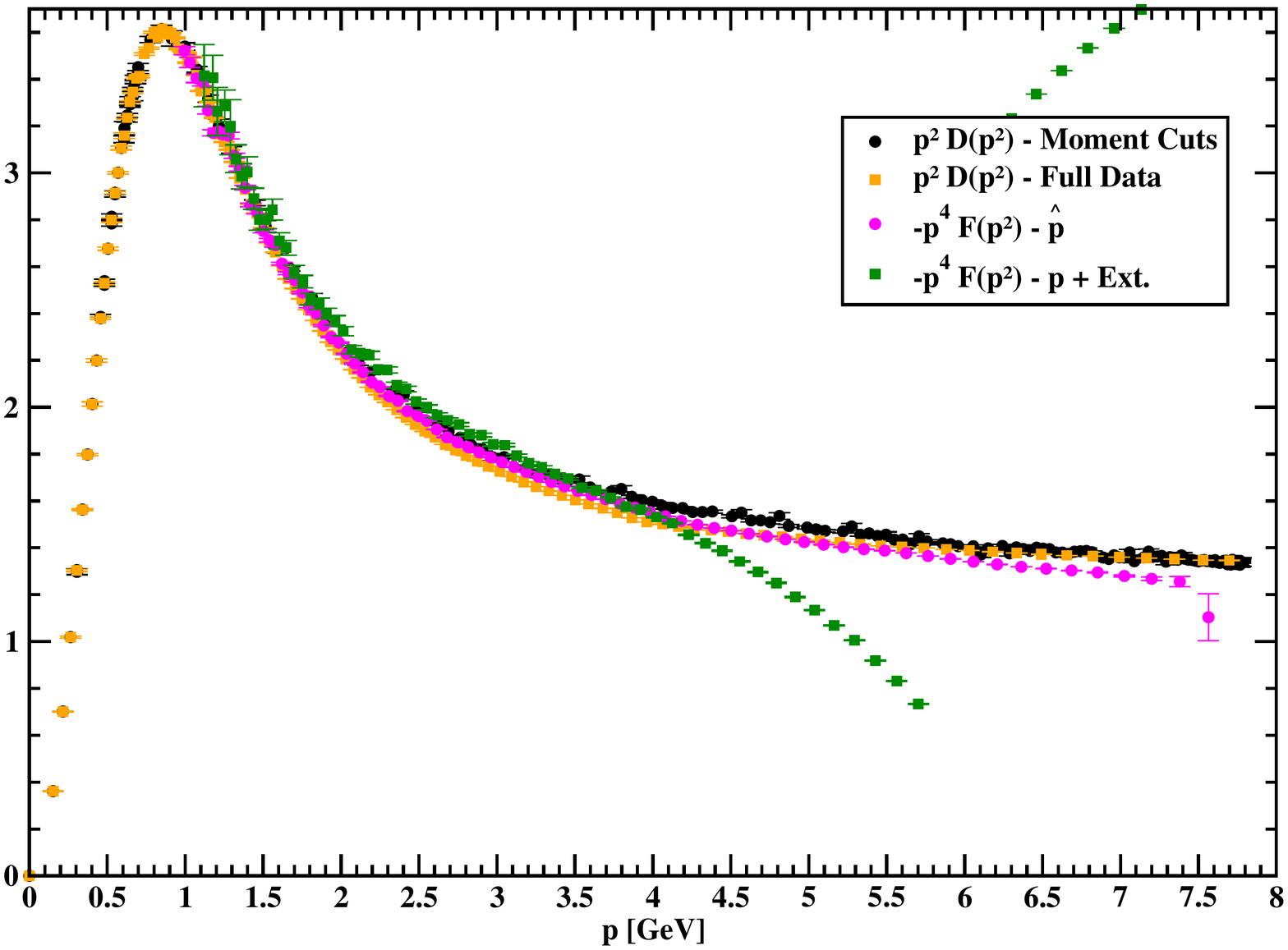} 
   \caption{The binned gluon dressing function for the form factors $p^2 E(p^2)$ (left) and $-p^4 F(p^2)$ (right) associated with the enlarged
                 basis (\ref{Eq:LandauLattProp}) as a function of the improved momentum $\hat{p}$ and of the naive momentum $p$ after $H(4)$  extrapolation.
                 Note that for $p \approx 5.8$ GeV onwards, the $H(4)$ extrapolation is no longer possible and the points reported do not take into
                 account the linear extrapolation in $p^{[4]}$. We call the reader attention to the magnitude of the correction.}
   \label{fig:FF_Cont_Basis_Binned_Comp}
\end{figure*}

The binned data, see Fig. \ref{fig:FF_All_Basis_Bin2p5},
 for the description that uses the improved lattice momentum results in a smooth curve that essentially follows the reference data 
reported in \cite{Dudal:2018cli}. However, the data in upper part of Fig. \ref{fig:FF_All_Basis_Bin2p5} shows also a consistent pattern 
where almost all the form factors are consistently below the moments cut evaluation of $D(p^2)$.
Indeed, a zoom of any of the subplots in Fig. \ref{fig:FF_All_Basis_Bin2p5}  shows a pattern for the binned data that follows the pattern observed for
the full momenta set given in Fig. \ref{fig:FF_Cont_Basis_Bin2p5}. The exception are the data points associated with
$-\hat{p}^4 F(\hat{p}^2)$ and $-\hat{p}^4 H(\hat{p}^2)$. Fig. \ref{fig:FF_Cont_Basis_Binned_Comp} illustrates this behaviour by zooming
the plots for $\hat{p}^2 E(\hat{p}^2)$ and $-\hat{p}^4 F(\hat{p}^2)$. A similar plot can be shown for $-\hat{p}^4 H(\hat{p}^2)$.
The data for $\hat{p}^2 E(\hat{p}^2)$ overlaps with the full momenta data set computed with the continuum tensor basis, while
the data for $-\hat{p}^4 F(\hat{p}^2)$ is slightly closer to the reference data, named momentum cuts in Fig. \ref{fig:FF_Cont_Basis_Binned_Comp}.

The description of the form factors in terms of the naive momentum, combined with the extrapolation, is closer to the reference
data in comparison with the  description in terms of the improved momentum. However, significant deviations occur for
$p \gtrsim 3$ GeV. Furthermore, a close look at the data in Fig. \ref{fig:FF_Cont_Basis_Binned_Comp}
shows that the extrapolated data is not on top of the reference data but, typically, exceeds it. Oncemore, it is for
$-\hat{p}^4 F(\hat{p}^2)$ that the agreement with $D(p^2)$ is better for momenta up $\sim 4$ GeV. 

For the continuum basis, the description of $D(p^2)$ in terms of the naive momentum combined with a linear extrapolation
are illustrated in Fig. \ref{fig:LatContBas80}. 
The linear extrapolation recovers the functional form seen in the reference gluon data. Note, however, that even after the $p^{[4]}$
extrapolation there are small discrepancies between the two data sets.

\section{Orthogonality and Completeness of the Tensor Basis \label{Sec:Orho}} 

\begin{figure}[t] 
   \centering
   \vspace{-1cm}
   \includegraphics[width=6in]{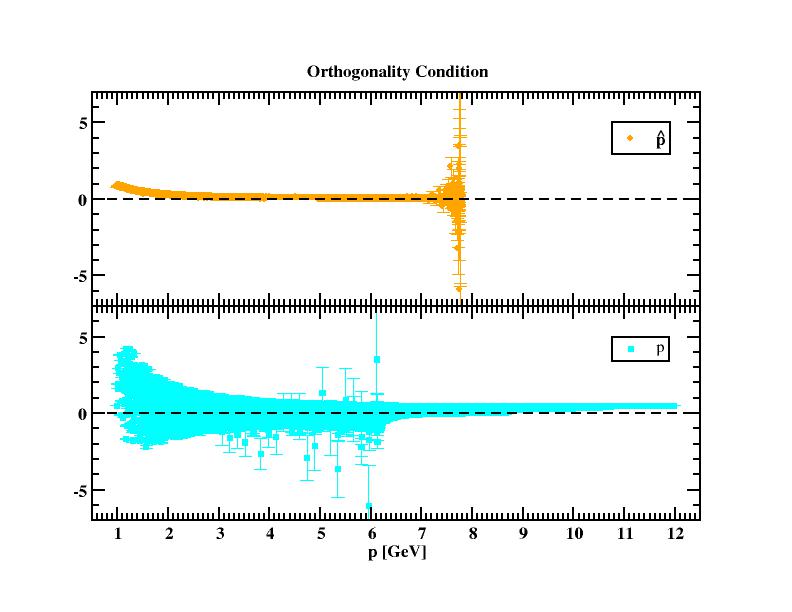}
   \caption{The orthogonality condition (\ref{Eq:TransvLatt}) for a general kinematical configurations written in terms of the improved momentum (upper plot)
                 and the lattice momentum (lower plot). For the lattice momentum, no $H(4)$ extrapolation was considered.}
   \label{fig:orthog}
\end{figure}

The orthogonality of the lattice gluon propagator was discussed in Sec. \ref{Sec:orto_gen} for a general kinematical configuration
and in Sec. \ref{Sec:SpecialKinematics} for the special kinematical configurations where the projection operators (\ref{Eq:Proj_E}) - (\ref{Eq:Proj_I}) 
are not all well defined.  
The so-called continuum relations between the form factors mentioned previously are a manifestation of the orthogonality of the
gluon field and, therefore, of the gluon propagator. In the continuum formulation, they are a consequence of the Slavnov-Taylor identity for the gluon. 

The orthogonality of the gluon propagator was partially studied in the previous section and we conclude that, in general, there is a range of momenta, 
that depends on the tensor basis, where the lattice gluon propagator is orthogonal and, therefore, the continuum Slavnov-Taylor identity is satisfied by
the lattice gluon propagator data.

The orthogonality condition, as given by Eq. (\ref{Eq:TransvLatt}), tests the relations between the various form factors measured in the Monte Carlo simulation. 
The relation (\ref{Eq:TransvLatt}), computed using the lattice form factors, is given in Fig. \ref{fig:orthog} for the tensor basis with the largest number of form
factors. The upper plot refers to the results when the improved momentum $\hat{p}$ is used, while the bottom plot
reports the results of computing the orthogonality condition with the lattice momentum $p$ and ignoring the $H(4)$ extrapolation. 
Note that the momenta in Fig. \ref{fig:orthog} do not include  diagonal and  on-axis momenta.

The data in Fig. \ref{fig:orthog} tests also the completeness of the tensor basis.
As discussed in Sec. \ref{Sec:GetD} (see text around Eq. (\ref{Eq:Aorto})) the replacement
of the standard definition of the gluon field by a definition where the gluon field is exactly orthogonal does not change the measured propagator.

As Fig.  \ref{fig:orthog} shows the gluon propagator form factors do not fulfil the relation given in Eq. (\ref{Eq:TransvLatt}) exactly and, in general, 
it is when the improved momentum is used to build the projectors that the above condition is better satisfied. 
As a test of the completeness of the tensor basis, these results suggest the range of momenta where one should expected the larger deviations
from the continuum result. The comparison of Fig. \ref{fig:orthog} with Figs. \ref{fig:D_E_F_H_L80}, \ref{fig:FF_All_Basis} and \ref{fig:FF_All_Basis_Bin2p5} shows
a consistent pattern of deviations that are certainly a manifestation of the lattice artefacts.

\section{Summary and Conclusion \label{Sec:ultima}}

In this work we have used tensor representations of the $H(4)$ symmetry group of the lattice formulation of QCD in the computation of the Landau gauge
gluon propagator. 
Clearly, probing this symmetry impacts on the quality of the Monte Carlo signal when compared to the $Z_4$ symmetry, used in previous simulations,
that selects a subset of $H(4)$ equivalent momenta. 

Our analysis of the lattice data describe the form factors associated with different tensor bases either as a function of the na\"{\i}ve lattice momentum
$p$ or of the improved momentum $\hat{p}$. 
The description of the form factors in terms of the lattice momenta shows severe lattice artefacts that can be solved, for a limited range of momenta,
by performing a linear extrapolation of the form factors in terms of the $H(4)$ invariants defined in Eq. (\ref{Eq:ScalarInvarianst}). 
This can be seen, for a particular class of momentum configurations and assuming that the continuum tensor structure is valid on the lattice, 
in Fig. \ref{fig:LatContBas80}. For the continuum tensor basis, the lattice data is better described in terms of $\hat{p}$,
where the gauge condition is fulfilled satisfied but whose lattice artefacts are resolved with the help of the momentum cuts,  see Fig. \ref{fig:FF_Cont_Basis_NoBin}.

The investigation of $R$, defined in Eq. (\ref{Eq:Def_R_Latt}) and that measures how well a given tensor basis describes the full lattice data, shows 
that the continuum tensor structure of the propagator does not apply to the lattice Landau gauge gluon propagator, see Figs. \ref{fig:R_LatContBas80} and 
\ref{fig:LatRecAll}. A faithful description of the lattice data requires $R \approx 1$ that, as seen in Fig. \ref{fig:LatRecAll}, is recovered only when the
data is described in terms of $p$ and after performing the extrapolation to $p^{[4]} = 0$. 
As seen  in Fig. \ref{fig:LatRecAll} the increase on the number of components of the tensor basis takes $R$ closer to unity but the approach to its optimal
value seems to be slow. This result motivated us to look at the definition of the gluon field and, in particular, at the orthogonality condition of 
the propagator.  Our analysis show that the deviations from $R = 1$ are not related to the definition of the gluon field and that imposing exact orthogonality
on the gluon field does not change neither $R$, neither the final outcome of the propagator.

The analysis of the form factors for the various tensor bases summarised in Figs. \ref{fig:G_and_I_Pimp80} to \ref{fig:FF_Cont_Basis_Binned_Comp} show
that the lattice form factors follow, in general, the expected behaviour derived from the gluon Slavnov-Taylor identity. Indeed, the various functions 
are compatible within one standard deviation and reproduce the reference data set, i.e. the lattice Landau gauge gluon propagator is described by a unique
form factor. Furthermore, by combining the different estimations of the form factors, taking into account also the uncertainties associated with the definition 
of the lattice spacing, one can yield a continuum form factor $D(p^2)$ and estimate a theoretical uncertainty in the final result, 
see Figs. \ref{fig:FF_All_Basis_Bin2p5} and \ref{fig:FF_Cont_Basis_Binned_Comp}. 
The data in these Figs. show that for the description of the lattice data in terms of $\hat{p}$, the best agreement with the reference data set, 
over a wide range of momenta, is achieved for the largest tensor basis and for $\hat{p}^4 F(\hat{p}^2)$ and $\hat{p}^4 H(\hat{p}^2)$.
On the other hand, a description of the lattice data in terms of the na\"{\i}ve momentum $p$, combined with the linear extrapolation in $p^{[4]}$, 
results on a set of form factors that have a better overlap with the reference data set when compared with the outcome of using the same tensor basis 
but written in terms of $\hat{p}$. However, due to the requirements of the extrapolation, the range of momenta accessed by the extrapolated
data is smaller, see, once more, Figs. \ref{fig:FF_All_Basis_Bin2p5} and \ref{fig:FF_Cont_Basis_Binned_Comp}.

In Fig. \ref{fig:FF_Cont_Basis_Binned_Comp} we do a detailed comparison of various estimations of continuum propagator using the largest tensor basis
and compare these form factors to the reference data set , which is now named ``$p^2 D(p^2)$ - Momentum Cuts''. 
For the data plotted in l.h.s, the form factor $E(\hat{p}^2)$ reproduces the full reference data set obtained using the continuum tensor structure but
the data is below the reference data. On the other hand, the data for $E(p^2)$ follows closely the reference data, although it is typically slightly above it.
However, the extrapolation produces unreliable results above $p \sim 4.5$ GeV ($a \, p \sim 2.3$). 
The r.h.s. of Fig. \ref{fig:FF_Cont_Basis_Binned_Comp} show that the data associated with $F$ are closer to the reference data, when compared with the data 
on the l.h.s. The data associated with $F(\hat{p}^2)$ is clearly closer to the reference data and the data associated to $F(p^2)$ deviates from the reference data 
earlier than the $E(p^2)$ data in the l.h.s. 
The matching between the various estimations of the form factors is not perfect, with both methods following closely the reference data set over a wide range of 
momenta. The difference in the estimated  form factors are a measure of the theoretical uncertainty in the continuum propagator.  Moreover,
the data also shows that the traditional approach to the computation of the gluon propagator, that is based on the use of the continuum tensor structure combined
with the use of the improved momentum and momentum cuts, provides a reliable estimation of the continuum limit for $D(p^2)$.

Finally, in the last section we perform and analysis of the orthogonality condition, whose results are summarised in Fig \ref{fig:orthog}, confirming that the condition
is better fulfilled when written in terms of the improved lattice momentum.

The analysis of the gluon propagator performed here considers a single lattice spacing that corresponds to $\beta = 6.0$. According to \cite{Oliveira:2012eh}
for this $\beta$ value or higher values and for sufficiently large volumes, that the authors claim to be larger than $\sim (6.5$ fm$)^4$, the finite volume and
finite lattice spacings are small. However, for smaller $\beta$, as those was used to access the infrared gluon propagator with the Wilson action,
the lattice spacing effects are sizeable and certainly an analysis based on the tensor representations, preferably combined with the extrapolations in $p^{[4]}$,
can introduce significant corrections to the standard approach. In what concerns the use of the extrapolation, unfortunately, it does not work at the extreme
momenta. In this case the corrections can only be computed by looking at different tensor bases. However, if the tensor basis considered has a large number
of operators, the calculation of the corresponding form factors is more involved and, in practise,  it is difficult to access the form factors with smaller ensembles
of configurations. An analysis of the form factors are performed here can provide, at most, an estimation of the error on the theoretical analysis.

\appendix

\section{Lorentz components of the gluon propagator \label{Prop:LatExtBasis}}

The Lorentz components of the lattice gluon propagator in the extended tensor basis (\ref{Eq:LandauLattProp}) are
\begin{tiny}
\begin{eqnarray}
& &
\left( D_{\mu\nu} (p) \right) =  \nonumber \\
& & \left(
  \begin{array}{c@{\hspace{0.3cm}}c@{\hspace{0.3cm}}c@{\hspace{0.3cm}}c}
     E(p^2)  + F(p^2) \, {p}^2_1 +  G(p^2) \,  {p}^4_1 & 
                      H(p^2) \,  {p}_1 \, {p}_2 + I(p^2) \,  {p}_1 \, {p}_2  \left(   {p}^2_1 +  {p}^2_2 \right) & 
                      H(p^2) \,  {p}_1 \, {p}_3 + I(p^2) \,  {p}_1 \, {p}_3  \left(   {p}^2_1 +  {p}^2_3 \right)  &
                      H(p^2) \,  {p}_1 \, {p}_4 + I(p^2) \,  {p}_1 \, {p}_4  \left(   {p}^2_1 +  {p}^2_4 \right)  \\
                      & & & \\
     H(p^2) \,  {p}_2 \, {p}_1 + I(p^2) \,  {p}_2 \, {p}_1  \left(   {p}^2_2 +  {p}^2_1 \right) &
                      E(p^2)  + F(p^2) \, {p}^2_2 +  G(p^2) \,  {p}^4_2 & 
                      H(p^2) \,  {p}_2 \, {p}_3 + I(p^2) \,  {p}_2 \, {p}_3  \left(   {p}^2_2 +  {p}^2_3 \right)  &
                      H(p^2) \,  {p}_2 \, {p}_4 + I(p^2) \,  {p}_2 \, {p}_4  \left(   {p}^2_2 +  {p}^2_4 \right)  \\
                      & & & \\
     H(p^2) \,  {p}_3 \, {p}_1 + I(p^2) \,  {p}_3 \, {p}_1  \left(   {p}^2_3 +  {p}^2_1 \right) &
                      H(p^2) \,  {p}_3 \, {p}_2 + I(p^2) \,  {p}_3 \, {p}_2  \left(   {p}^2_3 +  {p}^2_2 \right) &
                      E(p^2)  + F(p^2) \, {p}^2_3 +  G(p^2) \,  {p}^4_3 & 
                      H(p^2) \,  {p}_2 \, {p}_4 + I(p^2) \,  {p}_2 \, {p}_4  \left(   {p}^2_2 +  {p}^2_4 \right)  \\
                      & & & \\
     H(p^2) \,  {p}_4 \, {p}_1 + I(p^2) \,  {p}_4 \, {p}_1  \left(   {p}^2_4 +  {p}^2_1 \right) &
                      H(p^2) \,  {p}_4 \, {p}_2 + I(p^2) \,  {p}_4 \, {p}_2  \left(   {p}^2_4 +  {p}^2_2 \right) &
                      H(p^2) \,  {p}_4 \, {p}_3 + I(p^2) \,  {p}_4 \, {p}_3  \left(   {p}^2_4 +  {p}^2_3 \right) &
                      E(p^2)  + F(p^2) \, {p}^2_4 +  G(p^2) \,  {p}^4_4 
  \end{array}
  \right)  \ . \nonumber \\
   \label{PropLorentzStruct}
\end{eqnarray}
\end{tiny}

\section*{Acknowledgments}

This work was supported by national funds from FCT – Funda\c{c}\~ao para a Ci\^encia e a Tecnologia, I.P., within the projects 
UIDB/04564/2020 and UIDP/04564/2020
G.T.R.C acknowledges financial support  from FCT (Portugal) under the project UIDB/04564/2020. 
P. J. S. acknowledges financial support from FCT (Portugal) under Contract No. CEECIND/00488/2017.

This work was granted access to the HPC resources of the PDC Center for High Performance Computing at the KTH Royal Institute of Technology, Sweden,
made available within the Distributed European Computing Initiative by the PRACE-2IP, receiving funding from the European Community’s Seventh Framework 
Programme (FP7/2007-2013) under grand agreement no. RI-283493. The use of Lindgren has been provided under DECI-9 project COIMBRALATT. 
We acknowledge that the results of this research have been achieved using the PRACE-3IP project (FP7 RI312763) resource Sisu based in Finland at CSC. 
The use of Sisu has been provided under DECI-12 project COIMBRALATT2. We also acknowledge the Laboratory for Advanced Computing at the
University of Coimbra (\url{http://www.uc.pt/lca}) for providing access to the HPC resource Navigator.

\end{document}